%% file: thesis_main.tex
\documentclass[12pt]{extarticle} 

\usepackage{amsmath}
\usepackage{amssymb}
\usepackage{listings}
\usepackage{tikz}
\usepackage{float}
\usetikzlibrary{arrows, automata, positioning}
\makeatletter
\def\bbordermatrix#1{\begingroup \m@th
  \@tempdima 4.75\p@
  \setbox\z@\vbox{%
    \def\cr{\crcr\noalign{\kern2\p@\global\let\cr\endline}}%
    \ialign{$##$\hfil\kern2\p@\kern\@tempdima&\thinspace\hfil$##$\hfil
      &&\quad\hfil$##$\hfil\crcr
      \omit\strut\hfil\crcr\noalign{\kern-\baselineskip}%
      #1\crcr\omit\strut\cr}}%
  \setbox\tw@\vbox{\unvcopy\z@\global\setbox\@ne\lastbox}%
  \setbox\tw@\hbox{\unhbox\@ne\unskip\global\setbox\@ne\lastbox}%
  \setbox\tw@\hbox{$\kern\wd\@ne\kern-\@tempdima\left[\kern-\wd\@ne
    \global\setbox\@ne\vbox{\box\@ne\kern2\p@}%
    \vcenter{\kern-\ht\@ne\unvbox\z@\kern-\baselineskip}\,\right]$}%
  \null\;\vbox{\kern\ht\@ne\box\tw@}\endgroup}
\makeatother
\usepackage{theorem}

\theoremheaderfont{\itshape} {\theoremstyle{break}
} \theoremstyle{break}
 \theoremstyle{break}
\newtheorem{Thm}{Theorem}[section] {\theoremstyle{plain}
  \theorembodyfont{\rmfamily}  }
{\theoremstyle{plain}
  \theorembodyfont{\rmfamily}  \newtheorem{Def}{Definition}[section]}

\usepackage{authblk}

\title{FPRAS Approximation of the Matrix Permanent in Practice\thanks{This article is based on an MS thesis by the first author, submitted to Rice University on June 12, 2020 \cite{newman2020fpras}. Research partially supported by NSF Grant no. IIS-1527668.}}
\author{James E. Newman and Moshe Y. Vardi}
\affil{Computer Science Department, Rice University}

\date{}

\usepackage[scale=.85]{geometry}

\usepackage{fancyhdr}
\usepackage{lastpage}
\usepackage{hyperref}

\begin{document}

  \maketitle
  \thispagestyle{fancy}
\input{abstract}


\input{chapters}


\nocite{*}


\bibliographystyle{acm}
\bibliography{bibliography}

\end{document}

%% file: abstract.tex
\begin{abstract}
The matrix permanent belongs to the complexity class \#P-Complete. It is generally believed to be computationally infeasible for large problem sizes, and significant research has been done on approximation algorithms for the matrix permanent. We present an implementation and detailed runtime analysis of one such Markov Chain Monte Carlo (MCMC) based Fully Polynomial Randomized Approximation Scheme (FPRAS) for the matrix permanent which has previously only been described theoretically and with big-Oh runtime analysis. We demonstrate by analysis and experiment that the constant factors hidden by previous big-Oh analyses result in computational infeasibility.
\end{abstract}

%% file: chapters.tex

\section{Introduction}
\label{ch:Intro}

\subsection{Background and Motivation}
\label{sec:bg}

	The matrix permanent is a function on $n \times n$ square matrices defined similarly to the familiar matrix determinant. They differ only in a sign factor applied to alternating terms in the determinant.
	
	That small change has a dramatic result. While the matrix determinant can be efficiently computed in polynomial time, the fastest known algorithm for computing the matrix permanent is Ryser's Algorithm, requiring $O(n 2^n)$ operations \cite{ryser63}.
	
	The matrix permanent (with nonnegative entries) can also be framed as a graph problem. By treating the matrix as the adjacency matrix of a bipartite graph of $2n$ nodes, the matrix permanent is equivalent to counting the number of perfect matchings on the graph.
	
	This problem is an example of a \#P-Complete problem, a complexity class containing many interesting and useful problems. In fact, the matrix permanent is often taken as a representative example of the class \#P-Complete, not least in the paper which first formally defined the complexity classes \#P and \#P-Complete \cite{valiant79}.
	
	Problems in \#P-Complete are believed, in the general case, to be computationally infeasible, leading to research in heuristic and approximate solution methods. Because problems in \#P-Complete are interreducible in polynomial time, an efficient approximation algorithm for one \#P-Complete problem could potentially lead to efficient approximation algorithms for many other interesting \#P-Complete problems.
	
	Of particular interest are approximation algorithms which are guaranteed to run in polynomial time and, with a certain probability, to produce results within a certain error bound. When such an algorithm makes use of a secondary input of random bits, it is referred to as a fully polynomial randomized approximation scheme (FPRAS). 
	
	The 2004 publication of an FPRAS for the matrix permanent was a significant milestone in the study of approximation algorithms for \#P-Complete problems. The FPRAS was presented with a theoretical analysis proving the necessary conditions on the quality of the results as well as polynomial big-Oh runtime \cite{js04}.
	
	Although the existence of an FPRAS for the matrix permanent was proven, the FPRAS was never implemented. In fact, the original Jerrum, Sinclair, and Vigoda paper \cite{js04}, as well as later refinements \cite{annealing}, do not present a directly implementable algorithm. Critical parameters relating to the number of samples required by the algorithm and frequency of sampling are described in big-Oh terms sufficient for proving the existence of the FPRAS. Of course, in practice, an algorithm cannot collect a big-Oh number of samples.
	
\subsection{Problem Addressed}
\label{sec:prob}

	We explore two questions. First, whether the FPRAS for the matrix permanent actually leads to a computationally feasible approximation algorithm. Second, how the FPRAS for the matrix permanent actually performs in practice.

\subsection{Contributions}
\label{sec:contrib}
	
	The algorithms presented here are not original to this paper or its authors. Ryser's algorithm for the matrix permanent was published by H.J. Ryser in 1963 \cite{ryser63}. A theoretical description and big-Oh probabilistic analysis of the FPRAS for the matrix permanent was published by Jerrum, Sinclair, and Vigoda in 2004 \cite{js04}. An improved big-Oh analysis of a slightly modified version of the FPRAS was published by Bez\'akov\'a, \v{S}tefankovi\v{c}, Vazirani, and Vigoda in 2006 \cite{annealing}.
	
	The primary contributions of this work are a detailed (non-asymptotic) probabilistic analysis of the sampling requirements of the FPRAS, a working implementation of the FPRAS for the matrix permanent based on that probabilistic analysis, and an experimental evaluation of that implementation. We show both by analysis and by experiment that the FPRAS is not computationally feasible.
	
	Secondary contributions include minor improvements to the analysis of the FPRAS and methods of computing necessary values during execution, noted in Section \ref{ch:FPRAS}, and a proposed explanation for the large margin between the probabilistic analysis and practical performance of the FPRAS for small matrices.
	


\section{Preliminaries}
\label{ch:prelim}

\subsection{The Matrix Permanent}
\label{sec:perm}

\begin{Def}{Matrix Permanent}

Let $S_n$ be the set of all permutations over $\{0, 1, ... , n-1\}$. The matrix permanent of an $n\times n$ matrix is given by
$$per(A) = \sum_{\sigma\in S_n}\prod_{i=0}^{n-1} a_{i,\sigma(i)}$$	
\end{Def}
Informally, the matrix permanent is the sum over all the products of the permutations of the matrix entries.

\begin{Def}{Matrix Determinant}

The matrix permanent is similar to the more familiar matrix determinant. Both are sums over products of permutations of the matrix entries. For the determinant, each product of a permutation is also multiplied by the sign of the permutation. 
$$det(A) = \sum_{\sigma\in S_n}\prod_{i=0}^{n-1} sign(\sigma)a_{i,\sigma(i)}$$
\end{Def}

The naive computation of both the permanent and the determinant takes $O(n!)$ time. However, the presence of the $sign(\sigma)$ term in the determinant leads to efficient methods of computation. Using Gaussian elimination, the determinant can be calculated in $O(n^3)$ time.

The matrix permanent, on the other hand, has thus far resisted all efforts at finding an efficient method of computation. The best known general algorithm for the permanent is due to H. J. Ryser, and can be calculated in $O(m 2^m)$ time \cite{ryser63}.

Restricting our attention to matrices whose entries take values in $\{0,1\}$, the matrix permanent is equivalent to counting the number of permutations $\sigma\in S_n$ such that $a_{i,\sigma(i)}=1$ for all $i$. In other words, the permanent of an $n\times n$ $\{0,1\}$ matrix $A$ is equal to the number of unique ways one can select $n$ 1-valued entries such that exactly one entry is present in each row and exactly one entry is present in each column.

The $\{0,1\}$ matrix permanent can also be framed as a graph problem. The $n\times n$ matrix $A$ is used as the adjacency matrix of a bipartite graph of $2n$ vertices. Construct a bipartite graph $G=(U,V,E)$ where $|U|=n$ and $|V|=n$. The edges of the graph are assigned according to $E=\{(u_i,v_j) | A_{ij}=1\}$. This approach can be extended to apply to matrices of nonnegative entries following a standard transformation.

	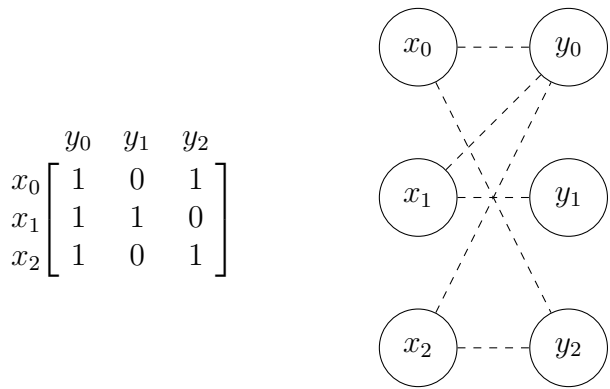
\begin{figure}[H]
		$$
		\vcenter{\hbox{\bbordermatrix{
			& y_0 & y_1 & y_2 \cr
				x_0 & 1 & 0 & 1 \cr
				x_1 & 1 & 1 & 0 \cr
				x_2 & 1 & 0 & 1	}}}
		\hspace{50pt}
		\vcenter{\hbox{
		\begin{tikzpicture}[node distance=2cm]
			\node[state] (x0) {$x_0$};
			\node[state] (x1) [below of=x0] {$x_1$};
			\node[state] (x2) [below of=x1] {$x_2$};
			\node[state] (y0) [right of=x0] {$y_0$};
			\node[state] (y1) [below of=y0] {$y_1$};
			\node[state] (y2) [below of=y1] {$y_2$};
			\draw[-, dashed] (x0) to (y0);
			\draw[-, dashed] (x0) to (y2);
			\draw[-, dashed] (x1) to (y0);
			\draw[-, dashed] (x1) to (y1);
			\draw[-, dashed] (x2) to (y0);
			\draw[-, dashed] (x2) to (y2);
		\end{tikzpicture}}}
		$$
	\caption{Adjacency matrix and corresponding bipartite graph with permanent 2.}
	\end{figure}
	
\begin{Def}{Perfect Matching}

On a bipartite graph $G=(U,V,E)$ with $|U|=|V|=n$, a \textit{perfect matching} is a set of edges $M\subseteq E$ with $|M|=n$ such that $\forall_{u\in U}\exists_{x\in V} (u,x)\in M$ and $\forall_{v\in V}\exists_{y\in U} (y,v)\in M$. Informally, a perfect matching is a set of edges in the graph such that every vertex in the graph is present in exactly one edge.
\end{Def}

A perfect matching on $G$ is equivalent to selecting $n$ 1-valued entries in the matrix such that exactly one entry is present in each row and exactly one entry is present in each column, as described above. Thus, computing the matrix permanent of an $n\times n$ $\{0,1\}$ matrix $A$ is equivalent to counting the number of perfect matchings in the bipartite graph $G=(U,V,E)$.
	
	\begin{figure}[H]
		$$
		\vcenter{\hbox{\bbordermatrix{
			& y_0 & y_1 & y_2 \cr
				x_0 & \textit{\textbf{1}} & 0 & 1 \cr
				x_1 & 1 & \textit{\textbf{1}} & 0 \cr
				x_2 & 1 & 0 & \textit{\textbf{1}}}}}
		\hspace{50pt}
		\vcenter{\hbox{
		\begin{tikzpicture}[node distance=2cm]
			\node[state] (x0) {$x_0$};
			\node[state] (x1) [below of=x0] {$x_1$};
			\node[state] (x2) [below of=x1] {$x_2$};
			\node[state] (y0) [right of=x0] {$y_0$};
			\node[state] (y1) [below of=y0] {$y_1$};
			\node[state] (y2) [below of=y1] {$y_2$};
			\draw[-,ultra thick] (x0) to (y0);
			\draw[-, dashed] (x0) to (y2);
			\draw[-, dashed] (x1) to (y0);
			\draw[-,ultra thick] (x1) to (y1);
			\draw[-, dashed] (x2) to (y0);
			\draw[-,ultra thick] (x2) to (y2);
		\end{tikzpicture}}}
		$$
	\caption{First 1-product permutation and corresponding perfect matching.}
	\end{figure}
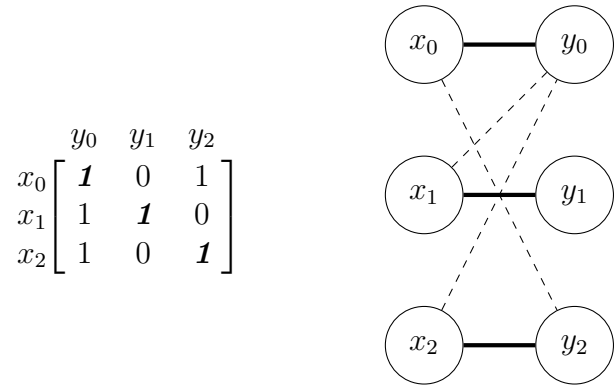

	\begin{figure}[H]
			$$
		\vcenter{\hbox{\bbordermatrix{
			& y_0 & y_1 & y_2 \cr
				x_0 & 1 & 0 & \textit{\textbf{1}} \cr
				x_1 & 1 & \textit{\textbf{1}} & 0 \cr
				x_2 & \textit{\textbf{1}} & 0 & 1	}}}
		\hspace{50pt}
		\vcenter{\hbox{
		\begin{tikzpicture}[node distance=2cm]
			\node[state] (x0) {$x_0$};
			\node[state] (x1) [below of=x0] {$x_1$};
			\node[state] (x2) [below of=x1] {$x_2$};
			\node[state] (y0) [right of=x0] {$y_0$};
			\node[state] (y1) [below of=y0] {$y_1$};
			\node[state] (y2) [below of=y1] {$y_2$};
			\draw[-, dashed] (x0) to (y0);
			\draw[-,ultra thick] (x0) to (y2);
			\draw[-, dashed] (x1) to (y0);
			\draw[-,ultra thick] (x1) to (y1);
			\draw[-,ultra thick] (x2) to (y0);
			\draw[-, dashed] (x2) to (y2);
		\end{tikzpicture}}}
		$$
	\caption{Second 1-product permutation and corresponding perfect matching.}	
	\end{figure}
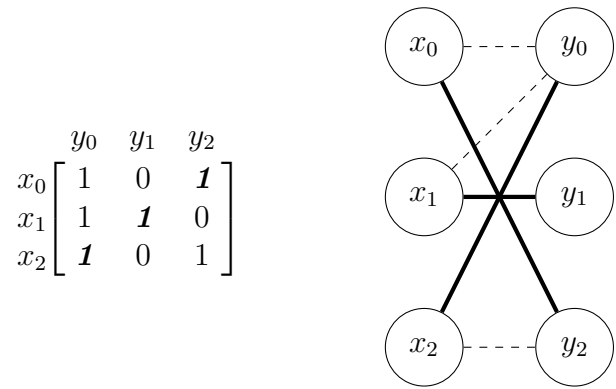

\subsection{Ryser's Algorithm}
\label{sec:ryser}

	The fastest known algorithm for the exact computation of the matrix permanent was published in 1963 by H. J. Ryser \cite{ryser63}. Ryser's algorithm uses the inclusion-exclusion principle to improve the computation time of the permanent.
	
	For an $n\times n$ matrix $A$, let $A_r$ with $0 < r < n$ denote the set of all matrices obtained from the matrix $A$ by replacing $r$ columns of $A$ with columns of exclusively 0-valued entries. The size of this set is $|A_r|= \genfrac(){0pt}{1}{n}{r} = \frac{n!}{r!(n-r)!}$. Also let $R(A) = \prod_{i=0}^{n-1}\sum_{j=0}^{n-1}a_{ij}$, that is, the product of the sums of the entries in each row.
	\begin{Def}{Ryser's Algorithm}
	
		For an $n\times n$ matrix $A$
			$$per(A) = R(A) - \sum_{A_1'\in A_1}R(A_1') + \sum_{A_2'\in A_2}R(A_2') - ... + (-1)^{n-1}\sum_{A_{n-1}'\in A_{n-1}}R(A_{n-1}')$$	
	\end{Def}
	
	The naive computation of this formula for the matrix permanent requires $n^2$ operations per application of $R$, and there are $O(2^n)$ applications of $R$, giving an overall computational time of $O(n^2 2^n)$. This can be reduced to $O(n 2^n)$ by computing each sum over $A_r$ in Gray code order, such that each subsequent $A_r'$ differs from its predecessor only in the position of a single zero-column. Thus, rather than fully computing each row sum ($O(n)$ operations per row) in each application of $R$, we add the single restored row entry and subtract the single removed row entry ($O(1)$ operations per row).
	
	Ryser's algorithm was implemented in Python to compute the exact matrix permanent to check the accuracy of the FPRAS implementation. For the purposes of our analysis, implementation of Ryser's algorithm in a more efficient programming language is unnecessary, as will become clear in the following section.

\subsection{P, NP, and \#P}
\label{sec:PNPSP}
	\begin{Def}{P}
	
		P is the class of decision problems (problems whose answer is either ``yes'' or ``no'', ``true'' or ``false'', or 1 or 0) which are solvable on a deterministic Turing machine in an amount of time which is a polynomial function of the size of the input. If the Turing machine accepts, the answer is ``yes'', ``true'', or 1, and if the Turing machine rejects, the answer is ``no'', ``false'', or 0 \cite{Cook71}.
	\end{Def}

	The decision problem version of bipartite perfect matchings can be phrased ``is there a perfect matching on the graph $G=(U,V,E)$?'' This problem is in P, that is, it can be decided in polynomial time whether a perfect matching exists for the graph $G=(U,V,E)$.

	\begin{Def}{NP}

		NP is the class of decision problems which are solvable on a nondeterministic Turing machine in an amount of time which is a polynomial function of the size of the input. Nondeterministic Turing machines can be characterized as Turing machines which, any time a choice is made, split into two copies of the machine, one copy having made one choice and the other copy having made the other choice. This results in the possibility of exponentially many computational branches. If any one computational branch accepts, the nondeterministic Turing machine accepts \cite{Cook71}.
	\end{Def}

	Obviously, $P\subseteq NP$ (and thus, bipartite perfect matchings are also in $NP$), but $P\stackrel{?}{=} NP$ is one of the most important open questions in complexity theory.
	
	\begin{Def}{NP-Complete}
		
		The set of problems in NP to which any problem in NP can be reduced in polynomial time is the complexity class NP-Complete \cite{Karp72}.
	\end{Def}
	
	The NP-Complete problems are the hardest problems in NP. No NP-Complete problem is known to be solvable in less than exponential time. Finding a polynomial-time solution to any NP-Complete problem would prove P=NP. If P$\neq$NP, then all NP-Complete problems are harder than all P problems. 

	\begin{Def}{\#P}
	
		For an NP problem, rather than deciding whether an accepting path exists in the nondeterministic Turing machine, we can count how many accepting paths exist. The set of counting problems corresponding to the decision problems in NP is the complexity class \#P \cite{valiant79}.
	\end{Def}
	
	Returning to the bipartite perfect matching problem, if the decision problem decides ``does a perfect matching exist in the graph $G=(U,V,E)$?'', the counting problem counts ``how many perfect matchings exist in the graph $G=(U,V,E)$?'' This is equivalent to computing the matrix permanent.
	
	\begin{Def}{\#P-Complete}
	
		The set of problems in \#P to which any problem in \#P can be reduced in polynomial time is the complexity class \#P-Complete \cite{valiant79}.
	\end{Def}

	It is generally trivial to prove that the counting problem associated with an NP-Complete decision problem is \#P-Complete. It is interesting to note that, although the bipartite perfect matching decision problem is in P, the bipartite perfect matching counting problem (the matrix permanent) is \#P-Complete. In fact, the complexity classes \#P and \#P-Complete were first described in a paper on the complexity of computing the matrix permanent \cite{valiant79}. A similar property holds for DNF satisfiability. Although DNF-SAT $\in$ P, \#DNF $\in$ \#P-Complete.
	
	As with NP-Complete problems, no \#P-Complete problem is known to be solvable in less than exponential time, and a polynomial-time algorithm for any \#P-Complete problem would prove P=NP.

\subsection{Approximation Algorithms}
\label{sec:approx}

	Because all known algorithms for solving \#P-Complete problems are worst-case exponential time, \#P-Complete problems may be computationally intractable for large or even moderate problem sizes. This has led to extensive research on approximation algorithms for \#P-Complete problems.
	
	\begin{Def}{$\epsilon - \delta$ Randomized Approximation Algorithms}
	
		An $\epsilon - \delta$ approximation algorithm for a problem with exact solution $X$ takes as an auxiliary input a stream of random bits and produces an approximate solution $\bar{X}$ such that $\Pr[\frac{X}{(1+\epsilon)}\leq \bar{X}\leq (1+\epsilon)X]\geq\delta$ with $0.5 < \delta < 1$ \cite{motwani}.
	\end{Def}
	
	$\epsilon$ is an input to the algorithm. $\delta$ may be an input to the algorithm, or may be a constant value, often $\frac{3}{4}$. A smaller value of $\epsilon$ (tighter bounds on the approximation) results in longer runtime. In the case of constant $\delta$, improved success probability $\delta^\prime$ can be obtained by performing a number of trials polynomial in $\frac{1}{1-\delta^\prime}$ and choosing the median value. 

	\begin{Def}{FPRAS}
	
		An $\epsilon-\delta$ randomized approximation algorithm which runs in time polynomial to the input, to $\epsilon$, and to $\delta$ is a fully polynomial randomized approximation scheme (FPRAS). An $\epsilon-\delta$ randomized approximation algorithm with constant $\delta$ which runs in time polynomial to the input and to $\epsilon$ is also an FPRAS because improved success probability $\delta^\prime$ can be obtained by performing a number of trials polynomial in $\frac{1}{1-\delta^\prime}$ and choosing the median value \cite{motwani}.

	\end{Def}
	
\subsection{Related Work}
\label{sec:related}
	Definitions of the complexity classes \#P and \#P-Complete were first published by Leslie Valiant in his 1979 paper ``The Complexity of Computing the Permanent'' \cite{valiant79}.
	
	Research on approximation algorithms for counting problems and their equivalents has a long history, but of particular interest is the 1986 paper ``Random Generation of Combinatorial Structures from a Uniform Distribution'' by Jerrum, Valiant, and Vazirani \cite{jvv86}. This paper formalizes the connection between uniform sampling and approximate counting as well as the connection between almost uniform sampling and randomized approximate counting. It proves that an FPRAS for a \#P problem exists if and only if an almost uniform sampling algorithm exists, and that the FPRAS and almost uniform sampling algorithm are inter-reducible \cite{jvv86}.
	
	The development of approximation algorithms for the matrix permanent has been a process of gradual improvement. An early paper ultimately published in 1993 describes a simple Monte Carlo approximation method for the matrix permanent which requires exponential time \cite{mcp93}.
	
	Due to the result in \cite{jvv86}, research into more efficient approximation algorithms for the matrix permanent has largely focused on improving the efficiency of almost uniform sampling algorithms. Andrei Broder's 1986 paper ``How Hard is it to Marry at Random? (On the Approximation of the Permanent)'' is an early example of an approximation algorithm using a Markov Chain Monte Carlo approach to compute the permanent in polynomial time for certain classes of matrices, but which requires exponential time in the general case \cite{broder86}. Jerrum and Sinclair's 1989 paper ``Approximating the Permanent,'' which referenced a preprint of \cite{mcp93}, is another example \cite{js89}.
	
	Refinement of that approach led to the FPRAS for the matrix permanent described in  Jerrum, Sinclair, and Vigoda's 2004 paper ``A Polynomial-Time Approximation Algorithm for the Permanent of a Matrix with Nonnegative Entries,'' and the improved analysis performed in Bez{\'a}kov{\'a}, {\v{S}}tefankovi{\v{c}}, Vazirani, and Vigoda's 2006 paper ``Accelerating Simulated Annealing for the Permanent and Combinatorial Counting Problems,'' which this paper analyzes \cite{js04,annealing}.
	
	A number of authors have explored randomized approximation algorithms for the matrix permanent which are not fully polynomial in the general case, but are of lower polynomial complexity than the FPRAS on certain classes of matrices \cite{apx1,apx2}. Another interesting result is a polynomial-time deterministic approximation algorithm for the matrix permanent of certain classes of matrices \cite{apx3}.

	However, at the time of publication of \cite{newman2020fpras}, the FPRAS described in \cite{js04} and \cite{annealing} remains the most efficient known FPRAS for the approximation of the matrix permanent.



\section{Analysis of the FPRAS}
\label{ch:FPRAS}

\subsection{Monte Carlo Method}
\label{sec:montecarlo}

	A common approach to estimating the value of a random variable is random sampling, also known as the Monte Carlo method. By taking some number of samples and averaging them, an estimate of the true value of random variable is produced. Standard probabilistic bounds can be applied to determine the sample size required to produce, with some probability, an estimate within some interval of the correct value \cite{motwani}.
	
	This approach can easily be applied to counting problems. In the case of the matrix permanent, for an $n \times n$ matrix there are $n!$ possible perfect matchings. Let $Z$ be a random variable representing the proportion of all possible perfect matchings which are actually present in the graph. By sampling the space of all possible perfect matchings, taking the value of a sample to be 1 if the sampled perfect matching is present in the graph and 0 if it is not and averaging over the number of samples, $\bar{Z}\approx Z$ can be estimated. It follows that $\bar{Z}*n!$ is an estimate of the count of perfect matchings present in the graph.
	
	One problem with this approach is that the required sample size depends on the true value of the random variable. For many problems, we have no a priori knowledge of the true value. As an illustration, consider a naive attempt to estimate the matrix permanent with sample set $S$ of polynomial size $|S|\leq n^b$, for some constant $b$, sampled uniformly from the space of possible perfect matchings. There are $n!$ possible perfect matchings. Consider an $n\times n$ matrix $A$ with a polynomial number of perfect matchings, $per(A)\leq n^c$ for some constant $c$. The probability of any individual sample $x$ being a perfect matching is $P(x=1) = \frac{n^c}{n!}$, and of not being a perfect matching, $P(x=0) = \frac{n!-n^c}{n!}$. For the set of samples $S$, the probability of sampling exactly zero perfect matchings is $P(\Sigma S = 0) =  (\frac{n!-n^c}{n!})^{n^b}$. In fact, for any chosen values of $b, c$, for large enough $n$, $P(\Sigma S = 0)$ becomes exponentially close to 1, and the chance of the sample set containing \textit{any} perfect matchings, $1 - P(\Sigma S = 0)$, becomes exponentially small. For matrices with polynomially many perfect matchings, if the matrix is large enough, the naive sampling algorithm will nearly always return an estimate of zero. 
	
	In fact, it is exactly this case--matrices with an exponentially small proportion of all possible perfect matchings present in the associated bipartite graph--for which many of the algorithms mentioned in Section \ref{sec:related} fail to approximate in polynomial time. 
	
	The key contribution of \cite{js04}, detailed in Section \ref{sec:mcperm}, is the construction of a Markov chain which results in an exponential increase in the probability of sampling a perfect matching from the graph when the proportion of perfect matchings present in the graph is exponentially small. Thus, a merely polynomial number of samples can provide a probabilistic guarantee of a useful sample set, regardless of the density or sparsity of the graph being sampled.
	
\subsection{Markov Chains}
\label{sec:markov}

	Constructing a Markov chain to sample from a distribution which has helpful properties not present in the uniform distribution is a common approach to improving Monte Carlo approximation methods.
	
	\begin{Def}{Markov Chain}
	
		A \textit{Markov Chain} is a discrete-time stochastic process consisting of a set of states S and a set of transition probabilities P such that $\forall{s_i, s_j \in S},  0\leq P_{ij} \leq 1$ and $\forall{s_i\in S},\sum_{s_j\in S}P_{ij}=1$.  $P_{ij}$ is the probability that, given the Markov Chain is in state $s_i$ at time $t$, the Markov Chain will be in the state $s_j$ at time $t+1$ \cite{motwani}.
	\end{Def}
	
	\begin{Def}{Irreducible Markov Chain}
		
		A Markov Chain is \textit{irreducible} if for every pair of states $s_i, s_j \in S$, a path exists from $s_i$ to $s_j$ and from $s_j$ to $s_i$ \cite{motwani}.
	\end{Def}
	
	\begin{Def}{Periodicity}
	
		The \textit{periodicity} of a state $s$ of a Markov Chain is the largest value $k$ such that, if $P[s^{(t)} = s] > 0$, for any $t' > t$, $P[s^{(t')}=s] > 0$ implies $t' = t + ck$ for some integer $c$. A state is \textit{aperiodic} if its periodicity is 1. A Markov Chain is \textit{aperiodic} if all states in the Markov Chain are aperiodic \cite{motwani}.
	\end{Def}
	
	\begin{Def}{Stationary Distribution}
		
		A \textit{stationary distribution} for a Markov Chain with transition probability matrix $P$ is a probability distribution $\pi$ on $S$ such that $\pi P = P$, that is, if the Markov Chain is in the stationary distribution at time $t$, it is also in the stationary distribution at time $t+1$ \cite{motwani}.
	\end{Def}

	\begin{Def}{Ergodic Markov Chain}
	
		A Markov Chain which is irreducible, aperiodic, and for which $|S|$ is finite is \textit{ergodic}. 	An ergodic Markov chain has a unique stationary distribution $\pi$ such that $\forall s\in S, \pi(s)>0$. Further, if $N(i,t)$ is the number of times state $i$ is visited in $t$ steps, $\forall s_i\in S, \lim\limits_{t\rightarrow\infty} N(i,t)=\pi(s_i)$, that is, after sufficiently many steps, an ergodic Markov Chain converges to its unique stationary distribution \cite{motwani}.
	\end{Def}

	These properties of Markov chains give rise to an obvious approach to solving our approximate counting problem. 
	
	We could construct an ergodic Markov chain with a state space that includes all possible perfect matchings. By construction, we could ensure that the Markov chain's unique stationary distribution assigns at worst a polynomially small sampling probability to the set of perfect matchings present in the graph (along with several other useful properties), and that the ratio between the probability assigned by the uniform distribution and by the stationary distribution is known precisely.
	
	Then we could use a polynomial number of samples to estimate the proportion of all possible perfect matchings present in the graph, and simply multiply this proportion by the probability distribution ratio (which we must somehow calculate) and the number of all possible perfect matchings (which is just $n!$).
	
	While the approach is obvious, how to actually construct such a Markov chain is not. Jerrum, Sinclair, and Vigoda succeeded in 2004. While their solution is broadly similar to the hypothetical Markov chain just discussed, it entails a great deal of additional complexity--almost entirely related to ensuring the proper stationary distribution is achieved--which will be described and analyzed in the next sections.
	
\subsection{Markov Chain for the Matrix Permanent \cite{annealing}}
\label{sec:mcperm}

	Jerrum, Sinclair, and Vigoda \cite{js04} developed a Markov Chain Monte Carlo approach, later improved by Bez\'akov\'a, \v{S}tefankovi\v{c}, Vazirani, and Vigoda \cite{annealing}, to produce a sampling distribution allowing a Monte Carlo-based FRPAS for the matrix permanent.
	
	The Markov Chain $\mathcal{M}$ used in the FPRAS has a state space consisting of all possible perfect matchings and all possible near-perfect matchings. The Markov Chain associated with the bipartite graph $G=(U,V,E)$ with $|U|=|V|=n$ has $|S|=(n^2+1)n!$, because each possible perfect matching has $n^2$ hole locations $(u,v)$. 
	
	Additionally, weights $w(u,v)$ are associated with each hole location and an activity variable $\lambda$ is used. For an edge $(u,v)$, we define $\lambda(u,v)=1$ if $(u,v)\in E$ and $\lambda(u,v)=\lambda$ otherwise. For a matching $M$, we define $\lambda(M)=\prod_{(u,v)\in M}\lambda(u,v)$ and $w(M) = \lambda(M)$ if $M$ is a perfect matching and $w(M) = w(u,v)\lambda(M)$ if $M$ is a near-perfect matching with hole at $(u,v)$. For a set of matchings $S'$, we define $\lambda(S')=\sum_{M\in S'} \lambda(M)$ and $w(S')=\sum_{M\in S'} w(M)$.
	
	Informally, the weight of a perfect matching is the product of the activities of the edges in the matching. The weight of a near-perfect matching is the product of the activities of the edges in the matching times the weight of the hole. By reducing the activity value in phases, the stationary distribution is biased towards matchings present in the graph, while the hole weights will be used to ensure that the sum of the probabilities of a set of matchings associated with a hole location are close to equal to the same sums for the other hole locations.
	
	Transitions $M_t\rightarrow M_{t+1}$ are made as follows:
	\begin{enumerate}
		\item If $M_t$ is a perfect matching, choose an edge $e$ uniformly at random from $M_t$ and let $M'=M_t\setminus e$.
		
		\item If $M_t$ is a near-perfect matching with hole at $(u,v)$, choose a vertex $x$ uniformly at random from $U\cup V$.
			\begin{enumerate}
				\item If $x\in (u,v)$, let $M' = M\cup (u,v)$.
				\item If $x\in V$ and $(w,x)\in M_t$, let $M'=M_t\cup (u,x)\setminus (w,x)$.
				\item If $x\in U$ and $(x,z)\in M_t$, let $M'=M_t\cup (x,v)\setminus (x,z)$.
			\end{enumerate}
			
		\item With probability $\min(1, \frac{w(M')}{w(M_t)})$, set $M_{t+1}=M'$, otherwise set $M_{t+1}=M_t$.
	\end{enumerate}

	Two key properties are proven about this Markov Chain in the paper.
	
	\begin{Thm}
		The stationary distribution of the Markov Chain is proportional to the weights of the matchings, that is, $\pi(M) = \frac{w(M)}{Z}$ where $Z = \sum_{M^\prime\in\Omega} w(M^\prime)$.
	\end{Thm}

	This property is essential. By manipulating the matching weights, we can manipulate the sample distribution such that potential matchings that are not present in the graph are exponentially unlikely to be sampled. Even if the count of perfect matchings present in the graph is exponentially small, we can approximate it with a polynomial number of samples.

	\begin{Thm}
		The Markov Chain is \textbf{rapidly mixing}, that is, the Markov Chain converges within total variation distance $\delta$ of the stationary distribution after a polynomial number of steps, in this case, $O(n^5log(n))$ steps.
	\end{Thm}
	
	This property is also essential. We wish to sample from the stationary distribution of the Markov Chain, but because the state space is exponentially large, the Markov chain might take exponentially many steps to converge to the stationary distribution, which would prevent our algorithm from running in polynomial time. 

\subsection{FPRAS for the Matrix Permanent \cite{annealing}}
\label{sec:fprasperm}

	The FPRAS for the Matrix Permanent proceeds in phases. Initially, all $w(u,v)=n$ and $\lambda=1$. In each phase, we reduce the activity $\lambda$, until finally, $\lambda = \frac{1}{n!}$, requiring $O(n^2log(n))$ phases. In each phase, we also update the weights to approximate the ideal weights for the current activity. The ideal weights are given by $w^*(u,v)=\frac{\lambda(\mathcal{P})}{\lambda(\mathcal{N}(u,v))}$. The weights are updated to approximate these ideal weights according to $w'(u,v)=\frac{\pi(\mathcal{P})}{\pi(\mathcal{N}(u,v))}w(u,v)$.
		
	Let $w_i(M)$ be the matching weight after the $i$th phase. $w_0(\Omega)=(n^2+1)n!$, and at termination, $w_r(\Omega)$ is approximately $(n^2+1)$ times the number of perfect matchings in $G$. In each phase, we can estimate $\frac{w_{i+1}(\Omega)}{w_i(\Omega)}$ by taking a set of $O(n^2log(n^4))$ samples $S'$ and computing  $\frac{w_{i+1}(S')}{w_i(S')}$. We can estimate $w_r(\Omega)$ by constructing the telescoping product: 
	$$w_0(\Omega) \times \frac{w_{1}(S'_1)}{w_0(S'_1)} \times \frac{w_{2}(S'_2)}{w_1(S'_2)} \times ... \times \frac{w_{r}(S'_r)}{w_{r-1}(S'_r)}$$
	We will let $\bar{Z_i} = \frac{w_{i+1}(S'_{i+1})}{w_i(S'_{i+1})}$. Thus, $w_r(\Omega)$ is approximately $(n^2+1)n!\bar{Z}_0\bar{Z}_1...\bar{Z}_{r-1}$. Observe that for any perfect matching $M\in\mathcal{M}_G$, $w(M)=1$. Then $w(\mathcal{M}_G)$ is equal to the number of perfect matchings on $G$.
	
	We can estimate $\frac{w_r(\mathcal{M}_G)}{w_r(\Omega)}$ by taking the sample mean of $\Theta(n^2\varepsilon^{-2})$ samples with value $1$ if the sample is in $\mathcal{M}_G$ and $0$ otherwise. Let $\bar{Y}$ denote this sample mean.
	
	Finally, we can approximate $w^*_r(\mathcal{M}_G)$ (the number of perfect matchings on $G$) by 
	\begin{equation}\label{eq:telescope}
		per(A) = w^*_r(\mathcal{M}_G) = (n^2+1)n!\bar{Z}_0\bar{Z}_1...\bar{Z}_{r-1}\bar{Y}
	\end{equation}
	The overall computation time required for this approximation is $O(n^7log^4(n))$.
	
	However, note that the input size in bits for this problem is $n' = n^2$, with the $n\times n$ matrix $A$ represented as a sequence of $n^2$ bits, so the runtime given the bitlength of the input is $O(n'^{3.5}log^4(n'))$.
	
\subsection{Analysis of Algorithm Parameters}
\label{sec:algoparam}

    The theoretical description of the FPRAS for the matrix permanent relies upon big-Oh analysis for simplicity. However, to implement the FPRAS in practice, we need at a minimum the constant factor associated with the big-Oh complexity. However, compared to an exact analysis, this formulation relies on increasing the constant factor associated with the highest order term to account for the lower order terms. If we wish to minimize runtime, we need as accurate an analysis as possible, including lower order terms. The full analysis of the algorithm parameters for the FPRAS is presented here, and comprises one of the primary contributions of this paper.
    
    \subsubsection{Useful Bounds for Approximation}
        Several bounds will be useful for this analysis.        
        \begin{equation} \label{eq:plustopow}
        	\forall x, 1+x\leq e^x	
        \end{equation}
		\begin{equation} \label{eq:powtotwice}
			x\in[0,1], e^x-1 \leq 2x	
		\end{equation}
		\begin{equation} \label{eq:powtooneplus}
			x\in[0,1], k\in \mathbb{N}^+, e^{x/(k+1)}\leq 1 + x/k
		\end{equation}
        \begin{equation} \label{eq:lnfacbound}
			x*\ln(x) - x + 1 \leq \ln(x!) \leq (x+1)\ln(x)-x+1
		\end{equation}
        \begin{Def}{Harmonic Number}
        
        The harmonic number $H_x$, the partial sum from $1$ to $x$ of the harmonic series, is given by $H_x = \sum_{j=1}^x \frac{1}{j}$. This sum is approximately equal to $\ln(x)$ with $\lim\limits_{x\rightarrow\infty}H_x = \ln(x) + \gamma$, where $\gamma \approx 0.577$ is the Euler--Mascheroni constant. 
        \end{Def}
        
        The harmonic number $H_x$ is bounded by:        
        \begin{equation} \label{eq:harmonicbound}
	        \ln(x) < \ln(x) + \frac{1}{x} \leq H_x \leq \ln(x) + 1
        \end{equation}

	\subsubsection{Probabilistic Inequalities}
	
		The analysis of the FPRAS makes use of two probabilistic inequalities. 
		
		\begin{Def}{Chebyshev's Inequality \cite{motwani}}
		
			For a random variable $X$, Chebyshev's inequality relates the standard deviation $\sigma$ or variance $Var(X)=\sigma ^2$ of a random variable $X$ to the proportion of samples expected to lie outside a given interval from the mean $\mu$.
			\begin{equation} \label{eq:chebyshev1}
				 \Pr(|X-\mu| \geq k\sigma) \leq \frac{1}{k^2}
			\end{equation}
			\begin{equation} \label{eq:chebyshev2}
				\Pr(|X-\mu| \geq \tilde{k}) \leq \frac{Var(X)}{\tilde{k}^2}
			\end{equation}
		\end{Def}
		
		\begin{Def}{Chernoff Bounds \cite{motwani}}
			
			Given the sum $\mathbf{X} = \sum X_i$ of $n$ independent Poisson trials on random variable $X$ with $\Pr(X_i = 1) = p_i$, $\mu = E(\mathbf{X}) = \sum p_i$, for any $\delta > 0$, the Chernoff bounds provide probabilistic bound on how far $\mathbf{X}$ will vary from $\mu$.
			\begin{equation} \label{eq:chernoff1}
				\Pr(\mathbf{X} > (1+\delta) \mu) < \Big(\frac{e^\delta}{(1+\delta)^{1+\delta}}\Big)^\mu
			\end{equation}
			\begin{equation} \label{eq:chernoff2}
				\Pr(\mathbf{X} < (1-\delta) \mu) < e^{-\frac{\mu\delta^2}{2}}
			\end{equation}
		\end{Def}
		
	\subsubsection{Markov Chain Mixing Time}
	
		The essential property of the Markov chain that enables this FPRAS is that it is rapidly mixing, that is, despite the exponential size of the Markov chain, it converges close to the stationary distribution in polynomial time. In other words, a representative sample set can be generated by a random walk through an exponentially small portion of the state space. We will reexamine this in the conclusion.

		The number of steps required to converge close to the stationary distribution is the \textit{mixing time} of the Markov chain. In particular, we will consider the mixing time necessary to converge within a certain total variation distance of the stationary distribution.
		
		The total variation distance between two distributions is defined as 
		\begin{equation} \label{eq:totalvariation}
			d_{TV}(\mu, \pi) = \frac{1}{2}\sum_{x\in \Omega}|\mu(x)-\pi(x)|=\sup_{x\in\Omega}|\mu(x)-\pi(x)|
		\end{equation}
		Informally, the total variation distance is the largest deviation between the actual distribution and the stationary distribution on any state in the Markov chain \cite{js04}.
		
		The mixing time to converge to variation distance of $\delta$ starting in state $x$ is given by 
		$$\tau_x(\delta) \leq \frac{7l\rho}{\pi(\mathcal{P})}(\ln(\frac{1}{\pi(x)}) + \ln(\frac{1}{\delta}))$$
		where $l$ and $\rho$ are properties of the Markov chain \cite{annealing}. The mixing time is proven using a canonical paths method, where $l$ is the maximum length of a canonical path and $l\leq n$, and $\rho$ is the congestion through states on the canonical path and $\rho \leq 12n$ \cite{annealing}. Further, we know that $\pi(\mathcal{P})\geq \frac{1}{4(n^2+1)}$ \cite{annealing}. Together, this leads to:
		\begin{equation*}
		\begin{split}
		\tau_x(\delta) &\leq 7*n*12n*4(n^2+1)(\ln(\frac{1}{\pi(x)}) + \ln(\frac{1}{\delta}))\\ 
		&= 336(n^4+n^2)(\ln(\frac{1}{\pi(x)}) + \ln(\frac{1}{\delta}))
		\end{split}
		\end{equation*}
		By choosing a ``good'' starting state--a perfect matching present in the graph at the beginning of the algorithm, and the last matching generated in the previous phase for later phases--we guarantee that 
		$$\pi(x) \geq \frac{1}{n!(n^2+1)}$$
		Thus,	
		\begin{equation} \label{eq:mixingtime}
		\begin{split}
		\tau(\delta) &\leq 336(n^4+n^2)(\ln(n!(n^2+1)) +\ln(\frac{1}{\delta}))\\
		&= 336(n^4+n^2)(\ln(n!)+\ln(n^2+1)+\ln(\frac{1}{\delta}))
		\end{split}
		\end{equation}
		
		However, once the stationary distribution has converged within total variation distance $\delta$, it is no longer necessary to wait the full mixing time between samples. In fact, the ``resampling'' time is simply the mixing time without the contribution of the probability of the start state \cite{bezakovaphd}:
		\begin{equation} \label{eq:resampletime}
			\tau_r(\delta) \leq 336(n^4+n^2)\ln(\frac{1}{\delta})
		\end{equation}
		
		The resampling time can also be derived from variations of the Chernoff bounds specific to Markov chain sampling rather than sets of independent samples \cite{aldous1987markov,lezaud1998chernoff}, based on an eigenvalue analysis of the Markov chain transition probabilities \cite{sinclair92}. This analysis is complex and produces a similar result.
		
		Defining an ``initialization'' time
		\begin{equation} \label{eq:inittime}
			\tau_i \leq 336(n^4+n^2)(\ln(n!) + \ln(n^2+1))
		\end{equation}
		we see that the convergence criteria are met in each phase by taking $\tau_i(\delta)$ initialization steps through the Markov chain once, then taking $\tau_r(\delta)$ resampling steps through the Markov chain before generating each sample.
		
		We also note that in the limit,
		\begin{equation} \label{eq:resampletimelim}
			\lim_{n\rightarrow\infty}\tau_r(\delta) \leq 336n^4\ln(\frac{1}{\delta})
		\end{equation}		
		
		and
		
		\begin{equation} \label{eq:inittimelim}
			\lim_{n\rightarrow\infty}\tau_i\leq 336n^4\ln(n!) = 336n^5\ln(n)	
		\end{equation}

	\subsubsection{Number of Phases to Approximate Weights}
		The number of phases required to approximate the weights is given as $O(n \times \ln(n))$. To compute the exact number of phases required, we examine the structure of the loops in the FPRAS algorithm:
		\begin{figure}[H]
		\begin{lstlisting}[mathescape, tabsize=4]
		let $\lambda_0=1$, $i=n$, $j=0$
			while $\lambda_j > \frac{1}{n!}$:
				$\lambda_{j+1} = 2^{-1/2i}\lambda_j$	
				if $i>2$ and $\lambda_{j+1} < (\frac{n}{n!})^{1/(i-1)}$:	
					$\lambda_{j+1} = (\frac{n}{n!})^{1/(i-1)}$
					$i = i - 1$
					$j = j + 1$
		return j
		\end{lstlisting}
		\caption{Algorithm for computing exact number of phases \cite{annealing}.}
		\label{fi:phases}
		\end{figure}
		
		\noindent
		First, for $i=n$, we repeatedly apply $\lambda_{j+1} = 2^{-1/2n}\lambda_j$ until $\lambda_{j+1} < (\frac{n}{n!})^{1/(n-1)}$.
		The number of phases required is:
		\begin{equation} \label{eq:phases1}
		 	\lceil \frac{2n}{(n-1)\ln(2)}log(\frac{n}{n!})\rceil
		 \end{equation}	
		The ceiling operation is required because we cannot perform a partial phase.
		
		At this point, $\lambda_{x+1} = (\frac{n}{n!})^{1/(n-1)}$ and $i = n-1$. Now, for each $i$ from $i = n-1$ to $i = 3$, the \textit{starting} value of $\lambda$ is $\lambda_{s} < (\frac{n}{n!})^{1/i}$ and we wish to reach an ending value of $\lambda$ of $\lambda_{f} < (\frac{n}{n!})^{1/(i-1)}$. This will take $y_i$ steps according to $\lambda_f = 2^{-y_i/2i}\lambda_s$, which gives $y_i = \lceil \frac{2}{(i-1)\ln(2)}\ln((n-1)!)\rceil$. This continues through $i=3$. The total phases required from $i=n-1$ to $i=3$ is:
		$$\sum_{i=3}^{n-1}\lceil \frac{2}{(i-1)\ln(2)}\ln((n-1)!)\rceil$$
		This can be simplified by shifting the index of the summation:
		\begin{equation} \label{eq:phases2}
			\sum_{i=2}^{n-2}\lceil \frac{2}{i*\ln(2)}\ln((n-1)!)\rceil
		\end{equation}
		After we reach $i = 2$, we have a starting value of $\lambda_s = (\frac{n}{n!})^{1/2}$ and wish to reach and ending value of $\lambda_f = \frac{1}{n!}$. The total phases required for $i=2$ is:
		\begin{equation} \label{eq:phases3}
			\lceil \frac{2}{\ln(2)}log(n!*n)\rceil
		\end{equation}
		Thus, combining \eqref{eq:phases1}, \eqref{eq:phases2}, and \eqref{eq:phases3}, the total number of phases in the algorithm is exactly:
		\begin{equation} \label{eq:exactphases}
		\begin{split}
			l &= \lceil \frac{2n}{(n-1)\ln(2)}\ln(\frac{n}{n!})\rceil\\
		 	&+ \left[\sum_{i=2}^{n-2}\lceil \frac{2}{i*\ln(2)}\ln((n-1)!)\rceil\right]\\
		 	&+ \lceil \frac{2}{\ln(2)}\ln(n!*n)\rceil
		\end{split}
		\end{equation}
		The ceiling operation is performed once for each $i$ with $n \geq i \geq 2$, so with 
		\begin{equation*}
		\begin{split}
			l^\prime &= \frac{2n}{(n-1)\ln(2)}\ln(\frac{n}{n!})\\
		 	&+ \left[\sum_{i=2}^{n-2} \frac{2}{i*\ln(2)}\ln((n-1)!)\right]\\
		 	&+ \frac{2}{\ln(2)}\ln(n!*n)
		\end{split}
		\end{equation*}
		we have $l^\prime\leq l \leq l^\prime + (n-1)$. We can simplify $l^\prime$ to:		
		\begin{equation}
			l^\prime = \frac{2}{\ln(2)} \bigg(\Big(\frac{n}{n-1}+H_{n-2}\Big)ln\Big((n-1)!\Big) + 2\ln(n)\bigg)
		\end{equation}		
		Making use of bounds \eqref{eq:harmonicbound} and \eqref{eq:lnfacbound} and by progressively loosening bounds on the result to eliminate lower order terms, for $n\geq 5$, we can derive the loose bounds:		
		\begin{equation}
			0.4 * n * \ln^2(n) \leq l \leq 7.22*n*\ln^2(n)
		\end{equation}
		We also note that in the limit,
		\begin{equation} \label{eq:phaseslim}
			\lim_{n\rightarrow\infty}l = \frac{2}{\ln(2)}H_{n}\ln(n!) = \frac{2}{\ln(2)}(\ln(n)n)\ln(n) = \frac{2}{\ln(2)}n * \ln^2(n)
		\end{equation}		
	
	\subsubsection{Samples per Phase to Approximate Weights}
	
		Given ideal weights, the distribution on the Markov chain would be \cite{js04}: 
		$$\hat{\pi}(\mathcal{P}) = \hat{\pi}(\mathcal{N}(u,v)) = \frac{1}{n^2+1}$$ 
		However, we only have estimates of the ideal weights. For ideal weights $w^*$, the estimated weights are bounded by $\frac{w^*}{2} \leq w \leq 2w^*$ \cite{annealing}. Thus, we know that the probability distribution of sampling the Markov chain within total variation distance $\delta$ is bounded by:		
		$$\pi(\mathcal{P}) , \pi(\mathcal{N}(u,v)) \geq \frac{1}{4(n^2+1)}-\delta$$		
		By choosing 
		\begin{equation} \label{eq:weightdelta}
			\delta \leq \frac{1}{8(n^2+1)}
		\end{equation}
		we ensure:		
		\begin{equation} \label{eq:minprob}
			\pi(\mathcal{P}) , \pi(\mathcal{N}(u,v)) \geq \frac{1}{8(n^2+1)}
		\end{equation}
		Further, we know that the ideal weights are related to the estimated weights by the relationship \cite{annealing}:	
		\begin{equation} \label{eq:weightupdate}
			w(u,v)^* = \frac{\pi(\mathcal{P})}{ \pi(\mathcal{N}(u,v))} w(u,v)
		\end{equation}
		
		We wish to estimate the values $\pi(\mathcal{P})$ and $\pi(\mathcal{N}(u,v))$ within a factor $2^{1/4}$ so that we can estimate the values $\frac{\pi(\mathcal{P})}{ \pi(\mathcal{N}(u,v))}$ within a factor $2^{1/2}$ \cite{annealing}. We can then use these estimates to improve our estimates of the ideal weights. For a set of perfect matchings $S$, let $S_p$ be the perfect matchings in $S$ and $S_{u,v}$ be the near-perfect matchings with a hole at $(u,v)$ in $S$. 
		
		For the perfect matchings, we have the Chernoff bounds:	
		$$\Pr\Big(|S_p| > 2^{1/4}E(|S_p|)\Big)< \Big(\frac{e^{(2^{1/4}-1)}}{(2^{1/4})^{(2^{1/4})}}\Big)^{E(|S_p|)}$$
		$$\Pr\Big(|S_p| < \frac{E(|S_p|)}{2^{1/4}}\Big) < e^{-E(|S_p|)(2^{1/4}-1)^2/2}=(e^{-(2^{1/4}-1)^2/2})^{E(|S_p|)}$$
		Because the estimate cannot be simultaneously too large and too small, we can combine the bounds by simply summing the failure probabilities:
		\begin{equation*}
		\begin{split}
		\Pr\Big(\frac{E(|S_p|)}{2^{1/4}} &\leq |S_p| \leq 2^{1/4}E(|S_p|)\Big)\\ &\geq 1 - \Big(\frac{e^{(2^{1/4}-1)}}{(2^{1/4})^{(2^{1/4})}}\Big)^{E(|S_p|)} - (e^{-(2^{1/4}-1)^2/2})^{E(|S_p|)}
		\end{split}
		\end{equation*}
		Because $\frac{e^{(2^{1/4}-1)}}{(2^{1/4})^{(2^{1/4})}} > e^{-(2^{1/4}-1)^2/2}$, we can simplify:
		\begin{equation} \label{eq:pbounds}
			\Pr\Big(\frac{E(|S_p|)}{2^{1/4}} \leq |S_p| \leq 2^{1/4}E(|S_p|)\Big)\geq 1 - 2\Big(\frac{e^{(2^{1/4}-1)}}{(2^{1/4})^{(2^{1/4})}}\Big)^{E(|S_p|)}
		\end{equation}
		An identical derivation produces a similar result for the near-perfect matchings.
		\begin{equation} \label{eq:nbounds}
			\Pr\Big(\frac{E(|S_{u,v}|)}{2^{1/4}} \leq |S_{u,v}| \leq 2^{1/4}E(|S_{u,v}|)\Big)\geq 1 - 2\Big(\frac{e^{(2^{1/4}-1)}}{(2^{1/4})^{(2^{1/4})}}\Big)^{E(|S_{u,v}|)}
		\end{equation}
		Because $E(\frac{|S_p|}{|S_{u,v}|}) = \frac{\pi(\mathcal{P})}{\pi(\mathcal{N}(u,v))}$, we can combine \eqref{eq:pbounds} and \eqref{eq:nbounds} into the bounds:
		\begin{equation*}
		\begin{split}
		Pr&\Big(\frac{1}{\sqrt{2}}\frac{\pi(\mathcal{P})}{\pi(\mathcal{N}(u,v))}
		\leq \frac{|S_p|}{|S_{u,v}|}
		\leq \sqrt{2}\frac{\pi(\mathcal{P})}{\pi(\mathcal{N}(u,v))}\Big)\\
		&\geq (1 - 2\Big(\frac{e^{(2^{1/4}-1)}}{(2^{1/4})^{(2^{1/4})}}\Big)^{E(|S_{p}|)})(1 - 2\Big(\frac{e^{(2^{1/4}-1)}}{(2^{1/4})^{(2^{1/4})}}\Big)^{E(|S_{u,v}|)})\\
		\end{split}
		\end{equation*}
		Because $\frac{e^{(2^{1/4}-1)}}{(2^{1/4})^{(2^{1/4})}} < 1$ and $E(|S_{p}|)> \frac{|S|}{8(n^2+1)}$ and $E(|S_{u,v}|)> \frac{|S|}{8(n^2+1)}$, we can simplify this to:
		\begin{equation*}
		\begin{split}
		\Pr\Big(\frac{1}{\sqrt{2}}\frac{\pi(\mathcal{P})}{\pi(\mathcal{N}(u,v))}
		&\leq \frac{|S_p|}{|S_{u,v}|}
		\leq \sqrt{2}\frac{\pi(\mathcal{P})}{\pi(\mathcal{N}(u,v))}\Big)\\
		&\geq 1 - 4\Big(\frac{e^{(2^{1/4}-1)}}{(2^{1/4})^{(2^{1/4})}}\Big)^\frac{|S|}{8(n^2+1)}
		\end{split}
		\end{equation*}
		
		This simplification ignores the small possibility that both values are outside their bounds by factors which cancel each other out, producing a result inside the combined bound. Because this small possibility increases the success probability, ignoring it results in a valid, slightly looser bound.
		
		However, this bound, of the form used in \cite{js04} and \cite{annealing},  significantly overestimates the required number of samples, not because of the simplification described in the previous paragraph, but because in each of $n^2$ estimates of $\frac{\pi(\mathcal{P})}{\pi(\mathcal{N}(u,v))}$, the probability of misestimating $S_p$ is counted.
		
		Instead of assigning a failure probability to the estimates of $\frac{\pi(\mathcal{P})}{\pi(\mathcal{N}(u,v))}$, we will assign failure probabilities to the estimates of $E(|S_p|)$ and $E(|S_{u,v}|)$. We wish to determine how many samples we must generate to generate an estimate within these bounds with failure probability $\hat{\eta}$:		
		$$\Pr\Big(\frac{E(|S_p|)}{2^{1/4}} \leq |S_p| \leq 2^{1/4}E(|S_p|)\Big)\geq 1 - \hat{\eta}$$
		$$\Pr\Big(\frac{E(|S_{u,v}|)}{2^{1/4}} \leq |S_{u,v}| \leq 2^{1/4}E(|S_{u,v}|)\Big)\geq 1 - \hat{\eta}$$
		Since the expressions for the failure probabilities of both bounds are identical, this entails:		
		$$\hat{\eta} = 2\Big(\frac{e^{(2^{1/4}-1)}}{(2^{1/4})^{(2^{1/4})}}\Big)^\frac{|S|}{8(n^2+1)}$$
		$$\frac{\hat{\eta}}{2} = \Big(\frac{e^{(2^{1/4}-1)}}{(2^{1/4})^{(2^{1/4})}}\Big)^\frac{|S|}{8(n^2+1)}$$
		$$\ln(\frac{\hat{\eta}}{2}) = \frac{|S|}{8(n^2+1)}\ln(\frac{e^{(2^{1/4}-1)}}{(2^{1/4})^{(2^{1/4})}})$$
		$$|S|=\frac{\ln(\hat{\eta}/2)8(n^2+1)}{\ln(\frac{e^{(2^{1/4}-1)}}{(2^{1/4})^{(2^{1/4})}})}
		= \frac{\ln(2/\hat{\eta})8(n^2+1)}{-\ln(\frac{e^{(2^{1/4}-1)}}{(2^{1/4})^{(2^{1/4})}})}$$
		\begin{equation} \label{eq:weightsampleseta}
		|S|\leq 475(n^2+1)\ln(2/\hat{\eta})
		\end{equation}
		
		Now, $\hat{\eta}$ is the failure probability of each of the $n^2 + 1$ estimates we must make per phase to update the weights, so the overall failure probability $\eta$ across all phases is
		\begin{equation} \label{eq:totalweightfailure}
			\eta = l (n^2+1) \hat{\eta}
		\end{equation}
		As we will see later, we will require the overall failure probability of this part of the algorithm to be no greater than $\frac{1}{12}$, yielding
		$$ \frac{1}{12} = l(n^2+1)\hat{\eta}$$
		\begin{equation}\label{eq:weightfailure}
			\hat{\eta} = \frac{1}{12l(n^2+1)}
		\end{equation}
		
		In the implementation of the algorithm, we will compute an exact value of $l$ to determine the required value of $\hat{\eta}$, but we can use the analytic bounds on $l$, the bound $\frac{0.9}{n^2} \leq \frac{1}{n^2+1}$, and the bound $n \geq 1.8 \ln^2(n)$ to produce loose bounds: 
		$$\hat{\eta} \geq \frac{1}{97n^3\ln^2(n)} \geq \frac{1}{54n^4}$$
		
		Combining \eqref{eq:weightsampleseta} and \eqref{eq:weightfailure} we generate a bound we will use to compute the number of samples per phase:
		\begin{equation} \label{eq:weightsamples}
			|S| \leq 475(n^2+1)\ln(2*12l(n^2+1))=475(n^2+1)\ln(24l(n^2+1))
		\end{equation}
		
		We can also generate a convenient, looser bound using the loose bound on $\hat{\eta}$, the bound $n^2+1 \leq 1.1n^2$, and the bound for $n>2$ that $2.1\ln(x) \geq (1.1706+\ln(x))$:
		\begin{equation} \label{weightsamplesloose}
		\begin{split}
			|S| &\leq 475*1.1 n^2 *\ln(108n^4) \leq 523n^2(\ln(108)+4\ln(n))\\
				&\leq 523n^2(4.6822 + 4\ln(n) \leq 2092n^2(1.1706+\ln(n))\\
				&\leq 2092n^2 2.1\ln(n) \leq 4394n^2\ln(n)
		\end{split}
		\end{equation}
		We also note that in the limit,
		\begin{equation}\label{weightsampleslim}
			\lim_{n\rightarrow\infty} |S| \leq 475n^2(\ln(24) + \ln(l) + \ln(n^2+1))=475n^2 2\ln(n)=950n^2\ln(n)
		\end{equation}
	
	\subsubsection{Samples per Phase for Approximate Counting}
		The weight approximation phases perform a dual purpose. In addition to producing an approximation to the ideal weights that result in the desired stationary distribution on the Markov chain, in each phase we produce a ratios $Z_i$ used in \eqref{eq:telescope}, the telescoping product expression for the approximate count of perfect matchings.
		
		From the bounds on the error in estimating the ideal weights and the rule for updating $\lambda$, for any matching $M$ we have:
		\begin{equation} \label{eq:ratioerror}
			\frac{1}{2} \leq \frac{w_{i+1}(M)}{w_i(M)} \leq 2	
		\end{equation}

		If we could sample exactly from the stationary distribution of the Markov chain, the expected mean of the sample set $S$ of matchings would be
		\begin{equation*}
			E(Z_i^\prime) = E(\frac{w_{i+1}(S)}{w_i(S)}) =	\frac{w_{i+1}(\Omega)}{w_i(\Omega)}
		\end{equation*}
		However, we can only sample from within variation distance $\delta$ of the stationary distribution. Given \eqref{eq:ratioerror}, the expected mean of the sample set $S$ must be within:
		\begin{equation*}
			(1-4\delta)\frac{w_{i+1}(\Omega)}{w_i(\Omega)}\leq E(Z_i)\leq (1+4\delta)\frac{w_{i+1}(\Omega)}{w_i(\Omega)}
		\end{equation*}
		If we sample the Markov chain within variation distance
		\begin{equation} \label{countingdeltaphase}
			\delta = \frac{\epsilon}{20l}	
		\end{equation}
		Then the expected mean is
		\begin{equation*}
			(1-\frac{\epsilon}{5l})\frac{w_{i+1}(\Omega)}{w_i(\Omega)}\leq E(Z_i)\leq (1+\frac{\epsilon}{5l})\frac{w_{i+1}(\Omega)}{w_i(\Omega)}
		\end{equation*}
		Using \eqref{eq:powtooneplus}, this becomes
		\begin{equation} \label{eq:ratiobound}
			e^{-\epsilon/4l}\frac{w_{i+1}(\Omega)}{w_i(\Omega)}\leq E(Z_i)\leq e^{\epsilon/4l}\frac{w_{i+1}(\Omega)}{w_i(\Omega)}
		\end{equation}
		Taking the product across all phases,
		\begin{equation} \label{eq:ratiobound1}
			e^{-\epsilon/4}\frac{w_{i+1}(\Omega)}{w_i(\Omega)}\leq E(Z_0Z_1...Z_{l-1})\leq e^{\epsilon/4}\frac{w_{i+1}(\Omega)}{w_i(\Omega)}
		\end{equation}
		Taking $|S|$ samples in each phase, from \eqref{eq:ratioerror},
		\begin{equation} \label{eq:phasevar}
			\frac{Var(Z_i)}{E(Z_i)^2} = \frac{(Z_i-E(Z_i))^2}{E(Z_i)^2} \leq \frac{9}{|S|}
		\end{equation}
		Across all phases, we have:
		\begin{equation} \label{eq:phasesvar}
			\frac{Var(Z_0Z_1...Z_{l-1})}{E(Z_0Z_1...Z_{l-1})^2} \leq \Big(1+\frac{9}{|S|}\Big)^l-1
		\end{equation}
		In \cite{js04}, approximations \eqref{eq:plustopow} and \eqref{eq:powtotwice} are used to simplify this expression to 
		\begin{equation} \label{eq:phasesvarjs}
			\frac{Var(Z_0Z_1...Z_{l-1})}{E(Z_0Z_1...Z_{l-1})^2} \leq 18l/|S|
		\end{equation}
		However, this is a substantially looser bound, so we will use \eqref{eq:phasesvar}. We bound the error in the partial telescoping product using Chebyshev's inequality:
		\begin{equation} \label{eq:zproderror}
		\begin{split}
			\Pr\Big(|Z_0Z_1...Z_{l-1}&-E(Z_0Z_1...Z_{l-1})| \geq \frac{\epsilon}{5}E(Z_0Z_1...Z_{l-1})\Big)\\ 
			&\leq \frac{5^2}{\epsilon^2}\frac{Var(Z_0Z_1...Z_{l-1})}{E(Z_0Z_1...Z_{l-1})^2}\\
			&\leq \frac{25}{\epsilon^2}\Bigg(\Big(1+\frac{9}{|S|}\Big)^l-1\Bigg)
		\end{split}
		\end{equation} \label{eq:zproderrorreq}
		As we will see later, our analysis requires
		\begin{equation}
			\Pr\Big(|Z_0Z_1...Z_{l-1}-E(Z_0Z_1...Z_{l-1})| \geq \frac{\epsilon}{5}E(Z_0Z_1...Z_{l-1})\Big) 
			\leq \frac{1}{12}
		\end{equation}
		Combining \eqref{eq:zproderror} and \eqref{eq:zproderrorreq}, we can solve for the required number of samples.
		\begin{equation*}
			\frac{25}{\epsilon^2}\Bigg(\Big(1+\frac{9}{|S|}\Big)^l-1\Bigg)\leq \frac{1}{12}
		\end{equation*}
		\begin{equation*}
			\Big(1+\frac{9}{|S|}\Big)^l-1\leq \frac{\epsilon^2}{300}
		\end{equation*}		
		\begin{equation*}
			\Big(1+\frac{9}{|S|}\Big)^l\leq \frac{\epsilon^2}{300}+1
		\end{equation*}
		\begin{equation*}
			1+\frac{9}{|S|}\leq \Big(\frac{\epsilon^2}{300}+1\Big)^{1/l}
		\end{equation*}
		\begin{equation*}
			\frac{9}{|S|}\leq \Big(\frac{\epsilon^2}{300}+1\Big)^{1/l}-1
		\end{equation*}
		\begin{equation} \label{eq:countingsamplesphases}
			|S|\geq \frac{9}{\big(\frac{\epsilon^2}{300}+1\big)^{1/l}-1}
		\end{equation}
		However, we will use \eqref{eq:phasesvarjs} to find a convenient bound for the required samples per phase for approximate counting in the limit (noting that even in the limit, this results in a looser bound than \eqref{eq:countingsamplesphases}).
		\begin{equation} \label{eq:countingsamplesphaseslim}
			\lim_{n\rightarrow\infty}|S| = \frac{5400 l}{\epsilon^2} = \frac{5400}{\epsilon^2}\frac{2}{\ln(2)}n * \ln^2(n)= \frac{10800}{\ln(2)}\frac{n*\ln^2(n)}{\epsilon^2}
		\end{equation}
		
	\subsubsection{Samples for Final Refinement of Approximate Count}
		Once the algorithm phases required for weight estimation and calculation of the partial telescoping product $Z_0Z_1...Z_{l-1}$ are complete, a final sampling step is required to refine the approximate count \cite{js04}. We generate a sample set $S^\prime$, taking each sample to have value $1$ if it is a perfect matching and $0$ if it is not, and take the sample mean, which we label $Y$.

		By taking $|S^\prime|$ samples from within variation distance of the stationary distribution
		\begin{equation}\label{eq:countdeltapost}
			\delta \leq \frac{\epsilon}{20}	
		\end{equation}
		we can follow the outline of the previous section to bound the expectation of $Y$ in terms of the stationary distribution of the set of all perfect matchings in the graph:
		\begin{equation}\label{eq:eybounds}
			e^{-\epsilon/4}\frac{w_{l}(\mathcal{M}_G)}{w_l(\Omega)}	\leq E(Y) \leq e^{\epsilon/4}\frac{w_{l}(\mathcal{M}_G)}{w_l(\Omega)}
		\end{equation}
		Because the samples only take values $0$ and $1$ and because $E(Y)\geq\frac{1}{4(n^2+1)}$, the variance of a single sample $Y^\prime$ is bounded by
		\begin{equation*}
			\frac{Var(Y^\prime)}{E(Y^\prime)^2}\leq \frac{1}{E(Y^\prime)}-1	\leq 4(n^2+1)-1 = 4n^2+3
		\end{equation*}
		The variance of the sample mean is bounded by
		\begin{equation}\label{eq:yvariance}
			\frac{Var(Y)}{E(Y)^2}\leq \frac{ 4n^2+3}{|S^\prime|}
		\end{equation}
		As in the last section, we will bound the error using Chebyshev's inequality.
		\begin{equation}\label{eq:ybound1}
			\Pr(|Y-E(Y)|\geq \frac{\epsilon}{5}E(Y))\leq \frac{25}{\epsilon^2}\frac{Var(Y)}{E(Y)^2} \leq \frac{25}{\epsilon^2}\frac{ 4n^2+3}{|S^\prime|}	= \frac{ 100n^2+75}{\epsilon^2|S^\prime|}
		\end{equation}
		We again require 
		\begin{equation}\label{eq:ybound2}
			\Pr(|Y-E(Y)|\geq \frac{\epsilon}{5}E(Y))\leq \frac{1}{12}
		\end{equation}
		Combining \eqref{eq:ybound1} and \eqref{eq:ybound2}, we solve for the required number of samples to refine the approximate count.
		\begin{equation*}
			\frac{ 100n^2+75}{\epsilon^2|S^\prime|} \leq \frac{1}{12}
		\end{equation*}
		\begin{equation}\label{countingsamplespost}
			|S^\prime|\geq \frac{1200n^2+900}{\epsilon^2}
		\end{equation}
		In the limit,
		\begin{equation}\label{countingsamplespostlim}
			\lim_{n\rightarrow\infty}|S^\prime|\geq \frac{1200n^2}{\epsilon^2}
		\end{equation}		
	
	\subsubsection{Summary of Algorithm Parameters}
	\label{sec:sumparam}
		Having derived the sampling parameters of the algorithm, we summarize them here. The algorithm consists of collecting sample sets during each of $l$ weight estimation phases and during a final refinement step. The exact number of weight estimation phases required is
		\begin{equation} \tag{\ref{eq:exactphases}}
		\begin{split}
			l &= \lceil \frac{2n}{(n-1)\ln(2)}\ln(\frac{n}{n!})\rceil\\
		 	&+ \left[\sum_{i=2}^{n-2}\lceil \frac{2}{i*\ln(2)}\ln((n-1)!)\rceil\right]\\
		 	&+ \lceil \frac{2}{\ln(2)}\ln(n!*n)\rceil
		\end{split}
		\end{equation}
		In the limit, the required number of weight estimation phases is
		\begin{equation} \tag{\ref{eq:phaseslim}}
			\lim_{n\rightarrow\infty}l = \frac{2}{\ln(2)}n * \ln^2(n)
		\end{equation}		
		Prior to collecting each sample set, we must take $\tau_i$ ``initialization steps.''
		\begin{equation} \tag{\ref{eq:inittime}}
			\tau_i(\delta) \leq 336(n^4+n^2)(\ln(n!) + \ln(n^2+1))
		\end{equation}
		Prior to collecting each sample in each sample set, we must take $\tau_r$ ``resampling steps.''
		\begin{equation} \tag{\ref{eq:resampletime}}
			\tau_r(\delta) \leq 336(n^4+n^2)\ln(\frac{1}{\delta})
		\end{equation}
		In the limit, this reduces to:
		\begin{equation} \tag{\ref{eq:resampletimelim}}
			\lim_{n\rightarrow\infty}\tau_r(\delta) \leq 336n^4\ln(\frac{1}{\delta})
		\end{equation}
		During the weight estimation phases, we have
		\begin{equation}
			\delta_w \leq \min(\frac{1}{8(n^2+1)}, \frac{\epsilon}{20l})	
		\end{equation}
		and the required number of samples is
		\begin{equation}
			|S_w| \geq \max(475(n^2+1)\\ln(24l(n^2+1)),\frac{9}{\big(\frac{\epsilon^2}{300}+1\big)^{1/l}-1})
		\end{equation}
		And in the limit (making use of the looser bound on samples required during phases for approximate counting),
		\begin{equation}\tag{\ref{weightsampleslim}}
			\lim_{n\rightarrow\infty} |S_w| =\max(950n^2\ln(n),\frac{10800}{\ln(2)}\frac{n*\ln^2(n)}{\epsilon^2})
		\end{equation}				
		During the final refinement step, we have
		\begin{equation}\tag{\ref{eq:countdeltapost}}
			\delta_c \leq \frac{\epsilon}{20}	
		\end{equation}
		and require 
		\begin{equation}\tag{\ref{countingsamplespost}}
			|S_c|\geq \frac{1200n^2+900}{\epsilon^2}
		\end{equation}
		In the limit,
		\begin{equation}\tag{\ref{countingsamplespostlim}}
			|S_c|\geq \frac{1200n^2}{\epsilon^2}
		\end{equation}		
		Thus, the total runtime of the weight estimation phases is:
		\begin{equation}
			T_w = (\tau_i(\delta_w) + \tau_r(\delta_w)|S_w| )l
		\end{equation}
		The fully expanded equation for the exact runtime, now completely derived, is too long to include as a single equation. However, we can consider its value in the limit.	 Because we will always take at least $O(n^2) > O(\ln(n!) + \ln(n^2+1))$ samples in a sample set, the total resampling steps taken during the collection of a sample set dominate the initialization steps. Using the expressions for the values in the limit, we find:
		\begin{equation}
		\begin{split}
			\lim_{n\rightarrow\infty}T_w = 
			&\frac{638400}{\ln(2)}n^6\ln^3(n)\\
			&*\max(2\ln(n), \ln(\frac{n}{\epsilon}))\\
			&*\max(n,\frac{10800}{950\ln(2)}\frac{\ln(n)}{\epsilon^2})\\
			\end{split}
		\end{equation}
		
		If $\epsilon \geq 1/n$, then $\max(2\ln(n), \ln(\frac{n}{\epsilon})) = 2\ln(n)$, otherwise $\max(2\ln(n), \ln(\frac{n}{\epsilon})) = \ln(\frac{n}{\epsilon})$.
		
		If $\epsilon \geq \sqrt{\frac{950\ln(2)n}{10800\ln(n)}}$ then $\max(n,\frac{10800}{950\ln(2)}\frac{\ln(n)}{\epsilon^2}) = n$, otherwise $\max(n,\frac{10800}{950\ln(2)}\frac{\ln(n)}{\epsilon^2}) = \frac{10800}{950\ln(2)}\frac{\ln(n)}{\epsilon^2}$.
		
		Thus, depending on the value of $\epsilon$, there are three values $\lim_{n\rightarrow\infty}T_w$ can take.
		
		\begin{equation}
		\lim_{n\rightarrow\infty}T_w = \begin{cases} 
      		\frac{7,257,600}{\ln^2(2)}n^6\ln^4(n)\ln(\frac{n}{\epsilon})\frac{1}{\epsilon^2}& \epsilon\leq 1/n \\
      		\frac{14,515,200}{\ln^2(2)}n^6\ln^5(n)\frac{1}{\epsilon^2} & 1/n < \epsilon < \sqrt{\frac{950\ln(2)n}{10800\ln(n)}}\\
      		\frac{1,276,800}{\ln(2)}n^7\ln^4(n) & \sqrt{\frac{950\ln(2)n}{10800\ln(n)}} \leq \epsilon
   		\end{cases}
		\end{equation}   
		
		The total runtime in the final refinement step is:
		\begin{equation}
		\begin{split}
		T_c &= \tau_i + \tau_r(\delta_c)|S_c| \\
		&= 336(n^4+n^2)(\ln(n!) + \ln(n^2+1)) + 336(n^4+n^2)\ln(\frac{20}{\epsilon})\frac{1200n^2+900}{\epsilon^2}
		\end{split}
		\end{equation}
		In the limit, this reduces to:
		\begin{equation}
		\lim_{n\rightarrow\infty}T_c = 403,200n^6\ln(\frac{20}{\epsilon})\frac{1}{\epsilon^2}
		\end{equation}
		Thus, in the limit, the total number of steps through the Markov chain is simply $\lim_{n\rightarrow\infty}T = \lim_{n\rightarrow\infty}T_w$.
		

\subsection{Analytic Computational Infeasibility}
\label{sec:ainfeas}

	Because the FPRAS is polynomial in the size of the matrix, while Ryser's algorithm is exponential in the size of the matrix, we know that no matter how large the constant factors associated with the asymptotic complexity of the FPRAS, there must be some $n^\prime$ such that for $n \geq n^\prime$, the required number of steps through the Markov chain is always less than the complexity of computing the exact permanent using Ryser's algorithm.
	
	Using the results from Section \ref{sec:algoparam}, we can easily compute the exact required number of steps through the Markov chain for any matrix size and error bound. Ignoring all other computation in the FPRAS, we can compare the required number of steps through the Markov chain to the complexity of computing the exact permanent using Ryser's algorithm.

	For small matrices, Ryser's algorithm is many orders of magnitude faster than the FPRAS. The smallest matrices in our experimental trials were of size $n=4$, for which Ryser's algorithm requires 64 operations, while the FPRAS (without relaxation of sampling parameters) requires 3,932,754,162,118 steps through the Markov chain.

	Indeed, comparing FPRAS runtime to Ryser's algorithm complexity over increasing $n$, we find that $n\geq 68$ is the point at which the FPRAS will require fewer steps through the Markov chain than the complexity of Ryser's algorithm. The number of steps through the Markov chain required for $n=68$ is 13,285,251,197,747,730,326,655, or approximately $1.33*10^{22}$. If we could perform one billion steps through the Markov chain per second, approximating the permanent of a $68\times 68$ matrix would take more than 420,984 years (using the Julian year of 365.25 days \cite{iaustyle}).
	
	The FPRAS for the matrix permanent is clearly computationally infeasible. 
	


\section{Experimental Evaluation}
\label{ch:experiment}

\subsection{Notes on Implementation}
\label{sec:impl}
	The implementation is largely a straightforward implementation of the algorithm as previously described in Sections \ref{sec:mcperm} and \ref{sec:fprasperm}, but some mathematically equivalent optimizations can be performed over the naive implementation of the algorithm. 

	For each Markov Chain step, we compute the probability $\min(1, \frac{w(M')}{w(M_t)})$ to decide whether to take the step or not. $w(M')$ has a factor $\lambda^i$ for $0\leq i\leq n$ and $w(M_t)$ has a factor $\lambda^j$ for $0\leq j\leq n$. Rather than computing $\lambda^i$ and $\lambda^j$ then performing division, we can replace $\frac{\lambda^i}{\lambda^j}$ by $\lambda^{i-j}$.
	
	For each phase, rather than storing the entire sample set for each phase, we can store, for perfect matchings and each hole position, for each number of edges with activity $\lambda$ rather than $1$, the number of samples found. This both reduces memory requirements and reduces the computation required to update the weights and compute $\bar{Z}$.
	
	Besides these optimizations, the sampling parameters (number of phases, initialization time, samples required per phase, resampling time during phases, samples required for the final refinement step, and resampling time for the final refinement step) must be precomputed using the results from Section \ref{sec:algoparam}.

\subsection{Parameter Relaxation}
\label{sec:relax}

	
	A final modification to the implemented algorithm is the ability to reduce the number of samples collected and the number of steps taken through the Markov chain between collecting samples below the requirements set by the theoretical analysis.
	
	Specifically, the notion of relaxation factors is introduced. Recall from Section \ref{sec:fprasperm} that the FPRAS consists of two overall stages. First, a sequence of weight estimation phases is performed. Then, a final approximation refinement stage is performed.

	Our implementation allows the specification of four independent relaxation factors. One relaxation factor, $r_{s,phase}$, applies to the number of samples collected during each weight estimation phase. One relaxation factor, $r_{t,phase}$, applies to the resampling interval during each weight estimation phase. One relaxation factor, $r_{s,final}$, applies to the number of samples collected during the approximation refinement stage. The final relaxation factor, $r_{t,final}$, applies to the resampling interval during the approximation refinement stage.
	
	In the FPRAS implementation, these relaxation factors serve as divisors of the respective sampling parameters required by the theoretical analysis. 
	
	Formally, if the number of samples is denoted $S$ and the resampling interval is denoted $T$, with the relaxed parameter indicated by a prime,
	
	$$S_{phase}^\prime = \frac{S_{phase}}{r_{s,phase}}$$
	
	$$T_{phase}^\prime = \frac{T_{phase}}{r_{t,phase}}$$
	
	$$S_{final}^\prime = \frac{S_{final}}{r_{s,final}}$$
	
	$$T_{final}^\prime = \frac{T_{final}}{r_{t,final}}$$
	
	For example, with $r_{s,phase}=10$ and $r_{t,phase}=2$, an approximation requiring 1000 samples per phase with a 500 step resampling time based on the theoretical analysis would instead perform the approximation using $1000/10$ samples per phase with a $500/2$ step resampling time. No other algorithmic parameters would be modified.
	
	The number of phases and initialization time used are always those required by the theoretical analysis. The number of phases can't be reduced without significantly altering the algorithm itself. The initialization time is executed once per phase, while the resampling time is executed for each sample in the phase, so the initialization time has a much smaller impact on runtime than the required number of samples and the resampling time.	
	
	Executing the algorithm with relaxed sampling parameters obviously invalidates the probabilistic proof of the failure rate and error bounds of the approximation. The implemented algorithm can certainly be used without parameter relaxation, however, due to the extreme computational inefficiency of the FPRAS, all trials were performed with some degree of parameter relaxation.
	
	However, many approximation algorithms perform better in practice than their proven theoretical guarantees, generally achieving much lower failure rates and smaller errors than specified. Given the option of parameter relaxation, we allow the possibility of extending experimental trials to larger matrices than would otherwise be practical, albeit with a presumably less accurate, more failure-prone approximation algorithm lacking theoretical guarantees on the results.
	
	Besides extending experimental trials to larger matrices than would otherwise be practical, parameter relaxation also gives us a potential avenue to explore how conservative the theoretical results are. By examining the performance of the relaxed approximation algorithm, we may be able to gain some insight into how much of a margin for possible future improvement exists between the sampling parameters required by the current theoretical analysis and the (hopefully smaller) sampling parameters required to produce good results.

\subsection{Experimental Setup}
\label{sec:setup}

	All experiments were performed on servers running Ubuntu 18.04.3 LTS with AMD EPYC 7501 CPUs. The Boost 1.65.1 library was used and the C++ code was compiled with gcc 6.5.0 with compilation flags ``-O3 -DNDEBUG''.
	
	For each $n\in\{4,6,8,10\}$, ten $\{0,1\}$ matrices were randomly generated with density $\frac{3}{4}n^2$ and ten $\{0,1\}$ matrices were randomly generated with density $\frac{7}{8}n^2$. The exact matrix permanent of each of these 80 matrices was computed using Ryser's Algorithm.
	
	The FPRAS implementation with $\epsilon = 0.5$ was then used to approximate the permanent for each matrix under different sampling parameter relaxations (including no relaxation when practical). For each trial, the estimate of the matrix permanent and the runtime of the trial in seconds were recorded. The algorithm can also fail completely, either by collecting $0$ samples for one hole position during one of the phases or by collecting $0$ samples of perfect matchings in the final refinement step. Because the algorithm can only produce positive estimates, in the case of this type of failure, the algorithm terminates and outputs an estimate of -1.
	
	Four rounds of experimental trails on the same set of 80 matrices were conducted. The discussion of our findings in the following sections will refer to small sets of trials relevant to the findings. The complete description of all experimental trials performed can be found in Appendix \ref{ch:appendix} along with their results.

\subsection{Experimental Computational Infeasibility}
\label{sec:einfeas}

	Using a Python program to generate the 80 random matrices (20 matrices for each $n\in \{4, 6, 8, 10\}$) used in all four rounds of trials, to compute the exact matrix permanent for each of the 80 matrices using Ryser's algorithm, and to write the 1,120 files defining the individual trials in the first round of testing took a cumulative total of 1.617 seconds.
	
	One set of trials from the second round is sufficient solid experimental evidence of the computational infeasibility of the FPRAS for the matrix permanent. In this set of trials, FPRAS estimation of the permanent of each of the 80 matrices was performed using $\frac{1}{256}$ the number of samples required to provide theoretical guarantees on the accuracy of the estimation and $\frac{1}{1024}$ the number of Markov chain steps between samples required to provide theoretical guarantees on the accuracy of estimation.
	
	Under these relaxation parameters, FPRAS estimation should require approximately $\frac{1}{262,144}$ its typical runtime. This decrease in runtime is not exact because the sampling parameters for the initial Markov chain warmup in each phase were not relaxed, but the number of steps in the warmup is a very small fraction of the number of steps taken in an entire phase.
	
	The average runtime in seconds \textbf{per matrix} for the exact computation of the matrix permanent with Ryser's algorithm as compared to the FPRAS estimation under the relaxation conditions just described are given below.
	
	\begin{table}[H]
	\begin{center}
		\begin{tabular}{| l | c | c | c | c | c |c | c | c | c |}
			\hline
			& $n=4$	& $n=6$	& $n=8$	& $n=10$\\ \hline
			FPRAS (relaxed) & 30 	& 738	& 10,528	& 50,634 \\ \hline
			Ryser's Algorithm & 0.0000377 & 0.000151 & 0.000661 & 0.00304 \\ \hline
		\end{tabular}
		\caption{Average runtime (in seconds) \textbf{per matrix}.}
		\label{ta:resruntime}
	\end{center}
	\end{table}
	
	Despite reducing the runtime of the FPRAS approximation by several orders of magnitude, approximating the matrix permanent for the 80 matrices whose exact permanents were computed in under two seconds required 344 \textbf{hours} of computation time.
	
	This set of trials also provides an ideal demonstration of why it was necessary to use parameter relaxation in all trials. Based on the total relaxation factor of 262,144 and the average approximation runtimes in these trials, we would expect unrelaxed FPRAS approximations for $n\in \{4,6,8,10\}$ to take on the order of three months, six years, ninety years, and four hundred years, respectively, of CPU time \textit{per matrix permanent approximated}.
	
\subsection{Approximation Accuracy}
\label{sec:acc}

	Besides demonstrating the computational infeasibility of the FPRAS for the matrix permanent, another important goal of implementing the FPRAS was to verify that the algorithm actually works, that is, that it produces accurate approximations.
	
	Given the demonstrated computational infeasibility of the FPRAS, it was simply not possible to perform FPRAS approximation of the permanent of a range of matrices of differing sizes and densities using the sampling parameters required by the theoretical analysis. 
	
	Upon consideration of the probabilistic inequalities upon which the theoretical analysis is based, it is obvious that collecting fewer samples than required or taking fewer steps between samples must increase the theoretical probability of failure compared to the required sampling parameters.
	
	Thus, if the FPRAS produces accurate approximations under relaxed sampling parameters, it is expected to produce accurate approximations under the theoretically required sampling parameters.
	
	To demonstrate that the FPRAS is correctly constructed and is capable of generating accurate approximations, we consider the fourth round of trials. One set of trials was performed with a phase resampling time relaxation factor of 33,554,432. This relaxation factor was chosen because it is larger than the theoretically required resampling time for any of the matrices used in the experimental trials, resulting in a resampling time of 1. In other words, samples were collected from consecutive steps through the Markov chain with no resampling delay at all. The number of required samples during the weight approximation phases was not relaxed. During the final refinement stage of the algorithm, the sample count was relaxed by a factor of 16 and the resampling time was relaxed by a factor of 64. 
	
	Despite the extreme degree of relaxation during the weight approximation phases and the moderate relaxation during the refinement stage, the average estimate error produced by the relaxed FPRAS across different matrix sizes ranged from 0.046 to 0.055, substantially beating the specified error bound of $\epsilon = 0.5$.
	
	While average error varied across the eight sets of relaxation parameters used in the fourth round of trials (totaling 640 individual approximations), none of the 640 trials exceeded the specified error bound, also substantially beating the allowed failure rate of $\frac{1}{4}$ despite sampling parameter relaxation. It is expected that the FPRAS without sampling parameter relaxation would perform at least as well on average across many problem instances. The proportional error in relaxed FPRAS estimation compared to the exact value computed with Ryser's algorithm and count of estimation errors (proportional error outside the specified $\epsilon=0.5$) are given below.
	
	\begin{table}[H]
	\begin{center}
		\begin{tabular}{| l | c | c | c | c |c | c | c | c |}
			\hline
			Relaxation Parameters & $n=4$	& $n=6$	& $n=8$	& $n=10$\\ \hline
			(1, 262,144, 16, 64) & 0.059 & 0.043 & 0.044 & 0.053 \\ \hline
			(1, 524,288, 16, 64) & 0.057 & 0.047 & 0.054 & 0.056 \\ \hline
			(1, 1,048,576, 16, 64) & 0.056 & 0.045 & 0.030 & 0.042 \\ \hline
			(1, 2,097,152, 16, 64) & 0.056 & 0.062 & 0.051 & 0.050 \\ \hline
			(1, 4,194,304, 16, 64) & 0.059 & 0.044 & 0.043 & 0.046 \\ \hline
			(1, 8,388,608, 16, 64) & 0.064 & 0.065 & 0.040 & 0.049 \\ \hline
			(1, 16,777,216, 16, 64) & 0.077 & 0.055 & 0.047 & 0.043 \\ \hline
			(1, 33,554,432, 16, 64) & 0.055 & 0.046 & 0.048 & 0.046 \\ \hline
		\end{tabular}
		\caption{Average proportional estimate error in fourth round of trials.}\label{ta:avgestimate4early}
	\end{center}
\end{table}
	
\begin{table}[H]
	\begin{center}
		\begin{tabular}{| l | c | c | c | c |c | c | c | c |}
			\hline
			Relaxation Parameters & $n=4$	& $n=6$	& $n=8$	& $n=10$\\ \hline
			(1, 262,144, 16, 64)* & 0 & 0 & 0 & 0 \\ \hline
			(1, 524,288, 16, 64) & 0 & 0 & 0 & 0 \\ \hline
			(1, 1,048,576, 16, 64) & 0 & 0 & 0 & 0 \\ \hline
			(1, 2,097,152, 16, 64) & 0 & 0 & 0 & 0 \\ \hline
			(1, 4,194,304, 16, 64) & 0 & 0 & 0 & 0 \\ \hline
			(1, 8,388,608, 16, 64) & 0 & 0 & 0 & 0 \\ \hline
			(1, 16,777,216, 16, 64) & 0 & 0 & 0 & 0 \\ \hline
			(1, 33,554,432, 16, 64) & 0 & 0 & 0 & 0 \\ \hline
		\end{tabular}
		\caption{Incidence of estimation outside specified bounds in fourth round of trials.}\label{ta:misestimates4early}
	\end{center}
\end{table}

	This is strong experimental evidence that the construction and analysis of the FPRAS is correct, and that it is an effective (if computationally infeasible) approximation algorithm.
	
\subsection{Untested Fundamental Concepts}
\label{sec:fundamental}
	
	There are two fundamental concepts that enable the FPRAS for the matrix permanent. 
	
	First, that when there are a ``small'' (polynomial) number of perfect matchings on the graph in the exponentially large space of possible perfect matchings, the exponentially few perfect matchings are assigned exponentially large weights in the sampling distribution of the Markov chain. This ensures that a polynomial number of samples can be used to accurately approximate a possibly exponentially small quantity, as described in Theorem 3.1 of Section \ref{sec:mcperm}.
	
	Second, that the Markov chain is rapidly mixing, as described in Theorem 3.2 of Section \ref{sec:mcperm}. Specifically, despite the exponential size of the Markov chain, the sampling distribution of the Markov chain is almost uniform after only polynomially many steps.

	Unfortunately, the computational infeasibility of the FPRAS renders us unable to demonstrate these properties in action. 
	
	In the first round of experimental trials, the results for $n=4$ and $n=6$ followed the expectations formed during manual testing in the process of implementing the FPRAS. The poor performance for $n=8$ and extremely poor performance for $n=10$ was unexpected. Under the hypothesis that the acceptable relaxation factors represent the margin between a conservative probabilistic analysis of the algorithm and its actual performance, we anticipated the acceptable relaxation factors would increase with $n$, and anticipated that some of the choices of relaxation factors used in the first round of trials would produce acceptable estimates for $n=8$ and $n=10$.

	To test whether we were simply increasing the relaxation factors too quickly, we undertook a second round of trials, in which we applied constant relaxation factors across $n\in\{4,6,8,10\}$. The second round of trials exposed a surprising result: relaxation parameter values that resulted in acceptable accuracy for $n=4$ and $n=6$ resulted in unacceptable accuracy for $n=8$ and $n=10$. This seems to contradict the previously mentioned hypothesis.

	We still believe that hypothesis to be true in the asymptotic case, that is, for large enough $n$, there are monotonically increasing functions of $n$ with value greater than 1 that are acceptable relaxation factors for the sampling parameters of the FPRAS, due to the conservative probabilistic analysis of the FPRAS.

	We propose that the observed decrease in acceptable relaxation factors over increasing $n$ results instead from a separate mechanism which greatly increases the acceptable relaxation factors for small $n$.

	Recall from Section \ref{sec:mcperm} that the state space of the Markov chain is factorially large. We require that the Markov chain be rapidly mixing because, asymptotically, in taking a polynomial number of samples with polynomial resampling time, we visit only an exponentially small fraction of the state space.

	However, as in the overall runtime analysis of Section \ref{sec:ainfeas}, for ``small'' values of $n$, a polynomial function of $n$ may be much larger than a factorial function of $n$. The size of the state space of the Markov chain is $(n^2+1)n!$. The number of states visited in one phase of the FPRAS is greater than the product of the required number of samples and the resampling time. Using the formulas in Section \ref{sec:sumparam}, we can compute the total number of steps taken through the Markov chain per phase. We can elide lower order terms to produce a simple, loose lower bound on the states visited per phase of $336n^4 * 475n^2 = 159,600 n^6$. Compare the exact values given $\epsilon = 0.5$ over the values of $n$ used in the trials:

\begin{table}[H]
	\begin{center}
		\begin{tabular}{| l | c | c | c | c |}
			\hline
							 & $n=4$	& $n=6$	& $n=8$	& $n=10$\\ \hline
			$|MC|$ & 408 & 26,640 & 2,620,800 & 366,508,800 \\ \hline
			$S_{phase}$ & 259,304 & 626,657 & 1,134,468 & 1,739,520 \\ \hline
			$S_{phase}/|MC|$ & 636 & 24 & 0.43 & 0.0047 \\ \hline
			$T_{phase}$ & $1.63*10^{11}$ & $2.17*10^{12}$ & $1.32*10^{13}$ & $5.18*10^{13}$ \\ \hline
			$T_{phase}/|MC|$ & 398,861,107 & 81,583,785 & 5,047,840 & 141,260 \\ \hline
		\end{tabular}
		\caption{Markov chain state space size compared to sample requirement and total steps per phase.}\label{ta:mcsizevsteps}
	\end{center}
\end{table}

In fact, the ratio $T_{phase}/|MC|$ drops below unity at $n=16$ and rapidly becomes exponentially small.

This comparison provides a compelling explanation for disproportionately large allowable relaxation factors for small $n$: the state space of the Markov chain is small enough that even with large relaxation factors, we take orders of magnitude more steps through the Markov chain than the size of the Markov chain itself. Essentially, for small $n$, a fundamental assumption of the probabilistic analysis of the FPRAS (that we visit only an exponentially small fraction of the space of the Markov chain) is violated in such a way that the FPRAS performs much better than the analysis would suggest.

The results of the third and fourth rounds of trials lend support to this proposed explanation. Phase sample requirement relaxation by a factor less than 128 resulted in acceptable approximation accuracy up to $n=10$ for every phase resampling time relaxation tested, even when the relaxation factor was large enough to eliminate the resampling delay completely, resulting in samples being produced by consecutive steps through the Markov chain.

For $n=4$ and $n=6$, both the phase sample requirement and the total number of steps are much larger than the size of the Markov chain. Even sampling consecutive steps through the Markov chain, we have many more samples than total states in the Markov chain, so we can reduce the sample requirement significantly without impacting the quality of the approximation. For $n=8$ and $n=10$, this is no longer the case: the number of samples per phase drops below the size of the Markov chain. As such, the quality of the result resumes its dependence on the probabilistic bounds on the variance of a ``small'' sample set, and we can no longer substantially relax the phase sample requirement. However, the number of samples is still close enough to the size of the state space of the Markov chain that even taking consecutive samples results in a representative sample set. As $n$ increases further, and the phase sample requirement becomes exponentially small compared to the size of the state space of the Markov chain, we expect this will cease to be the case and the allowable relaxation of the phase resampling time will be much more limited.

To summarize, for matrix sizes which are practical to actually test, both the polynomial resampling time and polynomial sample set size are of such high order and constant factor that they \textbf{do not} represent an exponentially small sample set taken during an exponentially short random walk through the Markov chain. The polynomial number of samples and polynomial path length are large compared to their respective spaces of exponential size. 

For large enough $n$, this would no longer be the case, but the computation time would be infeasibly long. Unfortunately, we are simply unable to demonstrate these critical concepts experimentally as part of the FPRAS for the matrix permanent.

\section{Conclusion}
\label{ch:conclusion}

Our results represent a mixed blessing. We demonstrated the overwhelming computational infeasibility of the ``fast'' polynomial-time FPRAS for the matrix permanent in Sections \ref{sec:einfeas} and \ref{sec:ainfeas}. 

In total, we performed 33 approximations, all under varying degrees of sampling relaxation (shortening the computation time relative to the FPRAS proper), for each of 80 matrices, totaling 2,640 experimental trials. Computing the exact matrix permanent for the 80 matrices took less than two seconds; computing it 33 times for each matrix (the same number of computations as our 2,640 trials) would take less than one minute. The 2,640 relaxed approximation trials performed required 6,947.62 hours (about 9.5 months) of CPU time.

We also noted the interesting phenomenon that the ``exponentially small'' sample space and random walk on the Markov chain, in fact, are not exponentially small for feasible problem sizes.

These are interesting results, but taken together, they prevent us from examining experimentally the behavior of the FPRAS when those critical properties are actually required and prevent us from being able to experimentally quantify the margin between the theoretical analysis of the sampling requirements of the FPRAS and the sampling requirements in practice. 

While the algorithm itself is not practically useful, it can serve as a useful example of the perils of asymptotic complexity analysis. In the conclusion of their paper describing their improved analysis of the FPRAS, Bez\'akov\'a et al. suggest that the existence of an $O(n^7log^4(n))$ FPRAS implies that ``computing permanents of $n\times n$ matrices with $n \approx 100$ now appears feasible.'' Returning to the analysis of Section \ref{sec:ainfeas}, under the same assumptions, the FPRAS approximation of the matrix permanent of a $100 \times 100$ matrix would require 64,022,847,298,779,435,144,166, or approximately $2.64*10^{23}$, steps through the Markov chain. Again assuming one billion steps per second, this approximation would require more than 8,366,379 years of computation time.

\appendix


\section{Complete Trial Description and Results}
\label{ch:appendix}

\subsection{Round 1}
For the initial round of trials, the base relaxation values of the parameters were chosen based on manual testing during implementation of the algorithm with $n\in \{3,4,5,6\}$. For $n=8$ and $n=10$, the initial relaxation values were chosen under the assumption that larger $n$ would tolerate larger relaxation values and based on the need to limit computation time. The base relaxations used for each $n$ are as follows:

\begin{table}[H]
	\begin{center}
		\begin{tabular}{| l | c | c | c | c |}
			\hline
			Parameter						& $n=4$	& $n=6$	& $n=8$	& $n=10$\\ \hline
			Samples per Phase 				& 64		& 256	& 512	& 1,024 	\\ \hline
			Resampling Time in Phases 		& 128	& 512	& 1,024	& 2,048	\\ \hline
			Samples for Refinement 			& 64		& 64	& 64		& 64 	\\ \hline
			Resampling Time for Refinement 	& 128	& 512	& 1,024	& 2,048	\\ \hline
		\end{tabular}
		\caption{Base relaxation factors for sampling parameters in first round of trials.}\label{ta:baserelax}
	\end{center}
\end{table}

Performance of the FPRAS under varying relaxation of the sampling parameters was evaluated by varying a single sampling parameter at a time by powers of 2 up to the final values (inclusive) as follows:

\begin{table}[H]
	\begin{center}
		\begin{tabular}{| l | c | c | c | c |}
			\hline
			Parameter						& $n=4$	& $n=6$	& $n=8$	& $n=10$\\ \hline
			Samples per Phase 				& 512	& 2,048	& 4,096	& 8,192	\\ \hline
			Resampling Time in Phases 		& 1,024	& 4,096	& 8,192	& 16,384	\\ \hline
			Samples for Refinement 			& 1,024	& 1,024	& 1,024	& 1,024	\\ \hline
			Resampling Time for Refinement 	& 1,024	& 4,096	& 8,192	& 16,384	\\ \hline
		\end{tabular}
		\caption{Maximum relaxation factors for sampling parameters in first round of trials.}\label{ta:maxrelax}
	\end{center}
\end{table}

There are five relaxation factors for samples for refinement and four relaxation factors for samples per phase and each resampling time. Thus there are 14 trials per matrix, resulting in a total of 1,120 trials. 

Despite each trial in the first round reducing the number of required steps through the Markov chain by a factor between 8,192 and 2,097,152, performing the initial round of 1,120 FPRAS approximation trials required 1506.84 hours of CPU time. 

\begin{table}[H]
	\begin{center}
		\begin{tabular}{| l | c | c | c | c |}
			\hline
			Relaxation Parameters & $n=4$	& $n=6$	& $n=8$	& $n=10$\\ \hline
			Base Case 		& 0.068 & 0.398 & 0.837 & 0.997 \\ \hline
			$2*r_{s, phase}$ 	& 0.081 & 0.637 & 0.975 & 1.000 \\ \hline
			$4*r_{s, phase}$ 	& 0.200 & 0.881 & 1.000 & - \\ \hline
			$8*r_{s, phase}$ 	& 0.326 & 0.993 & -     & - \\ \hline
			$2*r_{t, phase}$ 	& 0.081 & 0.387 & 0.824 & 0.997 \\ \hline
			$4*r_{t, phase}$ 	& 0.070 & 0.389 & 0.833 & 0.997 \\ \hline
			$8*r_{t, phase}$ 	& 0.090 & 0.380 & 0.828 & 0.997  \\ \hline
			$2*r_{s, final}$ 	& 0.157 & 0.389 & 0.836 & 0.997  \\ \hline
			$4*r_{s, final}$ 	& 0.204 & 0.425 & 0.829 & 0.997  \\ \hline
			$8*r_{s, final}$ 	& 0.214 & 0.355 & 0.827 & 0.997  \\ \hline			
			$16*r_{s, final}$ 	& 0.425 & 0.393 & 0.832 & 0.997  \\ \hline		
			$2*r_{t, final}$ 	& 0.088 & 0.397 & 0.835 & 0.997 \\ \hline
			$4*r_{t, final}$ 	& 0.108 & 0.366 & 0.827 & 0.997  \\ \hline
			$8*r_{t, final}$ 	& 0.097 & 0.411 & 0.829 & 0.997 \\ \hline
		\end{tabular}
		\caption{Average estimate error (excluding algorithm failures) across varying relaxation factors in initial trials.}\label{ta:avgestimate}
	\end{center}
\end{table}

\begin{table}[H]
	\begin{center}
		\begin{tabular}{| l | c | c | c | c |}
			\hline
			Relaxation Parameters & $n=4$	& $n=6$	& $n=8$	& $n=10$\\ \hline
			Base Case 	  & 0 & 1 & 20 & 20 \\ \hline
			$2*r_{s, phase}$ & 0 & 20 & 20 & 15$^*$ \\ \hline
			$4*r_{s, phase}$ & 0 & 20 & 19$^*$ & -$^*$ \\ \hline
			$8*r_{s, phase}$ & 1 & 20 & -$^*$ & -$^*$ \\ \hline
			$2*r_{t, phase}$ & 0 & 0 & 20 & 20 \\ \hline
			$4*r_{t, phase}$ & 0 & 2 & 20 & 20 \\ \hline
			$8*r_{t, phase}$ & 0 & 1 & 20 & 20  \\ \hline
			$2*r_{s, final}$ & 1 & 1 & 20 & 20  \\ \hline
			$4*r_{s, final}$ & 2 & 10 & 20 & 20  \\ \hline
			$8*r_{s, final}$ & 1 & 2 & 20 & 20  \\ \hline			
			$16*r_{s, final}$ & 6 & 7 & 20 & 20  \\ \hline		
			$2*r_{t, final}$ & 0 & 1 & 20 & 20 \\ \hline
			$4*r_{t, final}$ & 0 & 0 & 20 & 20  \\ \hline
			$8*r_{t, final}$ & 0 & 3 & 20 & 20 \\ \hline
		\end{tabular}
		\caption{Incidence of estimation outside specified bounds (excluding algorithm failures) across varying relaxation factors in initial trials. $^*$ indicates remaining trails were algorithm failures.}\label{ta:misestimates}
	\end{center}
\end{table}

\begin{table}[H]
	\begin{center}
		\begin{tabular}{| l | c | c | c | c |}
			\hline
			Relaxation Parameters & $n=4$	& $n=6$	& $n=8$	& $n=10$\\ \hline
			Base Case & 0 & 0 & 0 & 0 \\ \hline
			$2*r_{s, phase}$ & 0 & 0 & 0 & 5 \\ \hline
			$4*r_{s, phase}$ & 0 & 0 & 1 & 20 \\ \hline
			$8*r_{s, phase}$ & 0 & 0 & 20 & 20 \\ \hline
			$2*r_{t, phase}$ & 0 & 0 & 0 & 0 \\ \hline
			$4*r_{t, phase}$ & 0 & 0 & 0 & 0 \\ \hline
			$8*r_{t, phase}$ & 0 & 0 & 0 & 0  \\ \hline
			$2*r_{s, final}$ & 0 & 0 & 0 & 0  \\ \hline
			$4*r_{s, final}$ & 0 & 0 & 0 & 0  \\ \hline
			$8*r_{s, final}$ & 0 & 0 & 0 & 0  \\ \hline			
			$16*r_{s, final}$ & 0 & 0 & 0 & 0  \\ \hline		
			$2*r_{t, final}$ & 0 & 0 & 0 & 0 \\ \hline
			$4*r_{t, final}$ & 0 & 0 & 0 & 0  \\ \hline
			$8*r_{t, final}$ & 0 & 0 & 0 & 0 \\ \hline
		\end{tabular}
		\caption{Incidence of failure of algorithm across varying relaxation factors in initial trials.}\label{ta:failures}
	\end{center}
\end{table}

\begin{table}[H]
	\begin{center}
		\begin{tabular}{| l | c | c | c | c |}
			\hline
			Relaxation Parameters & $n=4$	& $n=6$	& $n=8$	& $n=10$\\ \hline
			Base Case 		& 687 	& 1,662	& 6,133	& 16,710 \\ \hline
			$2*r_{s, phase}$ 	& 341 	& 957	& 4,028	& 13,354 \\ \hline
			$4*r_{s, phase}$ 	& 179 	& 594	& 2,967	& - \\ \hline
			$8*r_{s, phase}$ 	& 98 	& 408	& -		& - \\ \hline
			$2*r_{t, phase}$ 	& 350 	& 968	& 4,083	& 13,391 \\ \hline
			$4*r_{t, phase}$ 	& 182 	& 591	& 3,028	& 11,679 \\ \hline
			$8*r_{t, phase}$ 	& 97 	& 417	& 2,523	& 10,914  \\ \hline
			$2*r_{s, final}$ 	& 679 	& 1,656 	& 6,176	& 16,944  \\ \hline
			$4*r_{s, final}$ 	& 681 	& 1,660	& 6,109	& 16,541  \\ \hline
			$8*r_{s, final}$ 	& 685 	& 1,650	& 6,158	& 16,602  \\ \hline			
			$16*r_{s, final}$ 	& 683 	& 1,660	& 6,087	& 16,506  \\ \hline		
			$2*r_{t, final}$ 	& 682 	& 1,667	& 6,134	& 16,649 \\ \hline
			$4*r_{t, final}$ 	& 682 	& 1,669	& 6,153	& 16,918  \\ \hline
			$8*r_{t, final}$ 	& 677 	& 1,667	& 6,192	& 16,509 \\ \hline
		\end{tabular}
		\caption{Average runtime (in seconds) across varying relaxation factors in initial trials (excluding algorithm failures.)}\label{ta:runtime}
	\end{center}
\end{table}

The base relaxation factors chosen in the first round of trials produced acceptable accuracy for $n=4$ and $n=6$, as did many of the variations in relaxation factors. However, for $n=8$ and $n=10$, every set of relaxation factors used in the first round of trials resulted in very poor accuracy. Further rounds of trials were performed to explore this behavior.

\subsection{Round 2}
Following the initial set of 1,120 trials, a second round of trials (on the same set of 80 randomly generated matrices) was run to evaluate the impact of varying the refinement stage relaxation factors. Each matrix was approximated using three sets of relaxation parameters. The three sets of relaxation parameters were the same for all values of $n$. The second round consisted of 240 additional trials.

\begin{table}[H]
	\begin{center}
		\begin{tabular}{| c | c | c | c |}
			\hline
			Samples 	& Resampling Time 	& Samples 			& Resampling Time \\
			per Phase 	& in Phases 			& for Refinement 	& for Refinement \\ \hline
			256			& 1,024				& 256				& 1,024 \\ \hline
			256			& 1,024				& 64					& 256 \\ \hline
			256			& 1,024				& 16					& 64 \\ \hline
		\end{tabular}
		\caption{Relaxation factors for second round of trials.}\label{ta:relax2}
	\end{center}
\end{table}


These 240 trials required an additional 1051.83 hours of CPU time.

\begin{table}[H]
	\begin{center}
		\begin{tabular}{| l | c | c | c | c |}
			\hline
			Relaxation Parameters & $n=4$	& $n=6$	& $n=8$	& $n=10$\\ \hline
			(256, 1,024, 256, 1,024) 	& 0.182 & 0.358 & 0.543 & 0.743 \\ \hline
			(256, 1,024, 64, 256) 	& 0.172 & 0.401 & 0.575 & 0.743 \\ \hline
			(256, 1,024, 16, 64) 	& 0.153 & 0.386 & 0.582 & 0.744 \\ \hline

		\end{tabular}
		\caption{Average estimate error in second round of trials.}\label{ta:avgestimate2}
	\end{center}
\end{table}

\begin{table}[H]
	\begin{center}
		\begin{tabular}{| l | c | c | c | c |c | c | c | c |}
			\hline
			Relaxation Parameters & $n=4$	& $n=6$	& $n=8$	& $n=10$\\ \hline
			(256, 1,024, 256, 1,024) 	  & 1 & 3 & 16 & 20 \\ \hline
			(256, 1,024, 64, 256) 	& 0 & 0 & 19 & 20 \\ \hline
			(256, 1,024, 16, 64) 	& 0 & 0 & 20 & 20 \\ \hline
		\end{tabular}
		\caption{Incidence of estimation outside specified bounds in second round of trials.}\label{ta:misestimates2}
	\end{center}
\end{table}

\begin{table}[H]
	\begin{center}
		\begin{tabular}{| l | c | c | c | c |c | c | c | c |}
			\hline
			Relaxation Parameters & $n=4$	& $n=6$	& $n=8$	& $n=10$\\ \hline
			(256, 1,024, 256, 1,024) & 0 & 0 & 0 & 0 \\ \hline
			(256, 1,024, 64, 256) 	& 0 & 0 & 0 & 0 \\ \hline
			(256, 1,024, 16, 64) 	& 0 & 0 & 0 & 0 \\ \hline
		\end{tabular}
		\caption{Incidence of failure of algorithm in second round of trials.}\label{ta:failures2}
	\end{center}
\end{table}

\begin{table}[H]
	\begin{center}
		\begin{tabular}{| l | c | c | c | c |c | c | c | c |}
			\hline
			Relaxation Parameters & $n=4$	& $n=6$	& $n=8$	& $n=10$\\ \hline
			(256, 1,024, 256, 1,024) 	& 30 	& 738	& 10,528	& 50,634 \\ \hline
			(256, 1,024, 64, 256) 	& 33 & 961 & 10,197 & 50,896 \\ \hline
			(256, 1,024, 16, 64) 	& 50 & 1,139 & 11,014 & 53,109 \\ \hline
		\end{tabular}
		\caption{Average runtime (in seconds) in second round of trials.}\label{ta:runtime2}
	\end{center}
\end{table}

The surprising result that constant relaxation factors resulted in worse performance for larger $n$ led to the discussion found in Section \ref{sec:fundamental}.

\subsection{Round 3}
A third round of trials (again, on the same set of 80 randomly generated matrices) was run to evaluate the relative impact of the sample quantity and resampling time relaxation factors. The product of the relaxation factors for the phases was kept constant (and the same as in the second round of trials). The relaxation factors for the refinement stage were kept constant at the smallest values used in the second round of trials. The third round consisted of 720 additional trials.

\begin{table}[H]
	\begin{center}
		\begin{tabular}{| c | c | c | c |}
			\hline
			Samples 	& Resampling Time 	& Samples 			& Resampling Time \\
			per Phase 	& in Phases 			& for Refinement 	& for Refinement \\ \hline
			512			& 512				& 16					& 64 \\ \hline
			256$^*$		& 1,024$^*$			& 16$^*$				& 64$^*$ \\ \hline
			128			& 2,048				& 16					& 64 \\ \hline
			64			& 4,096				& 16					& 64 \\ \hline
			32			& 8,192				& 16					& 64 \\ \hline
			16			& 16	,384				& 16					& 64 \\ \hline
			8			& 32,768				& 16					& 64 \\ \hline
			4			& 65,536				& 16					& 64 \\ \hline
			2			& 131,072			& 16					& 64 \\ \hline
			1			& 262,144			& 16					& 64 \\ \hline
			
		\end{tabular}
		\caption{Relaxation factors for third round of trials. $^*$ indicates trials retained from second round.}\label{ta:relax3-1}
	\end{center}
\end{table}

The 720 new trials in the third round required 3439.60 hours of CPU time.

\begin{table}[H]
	\begin{center}
		\begin{tabular}{| l | c | c | c | c |c | c | c | c |}
			\hline
			Relaxation Parameters & $n=4$	& $n=6$	& $n=8$	& $n=10$\\ \hline
			(512, 512, 16, 64) 	& 0.335 & 0.636 & 0.832 & 0.938 \\ \hline
			(256, 1,024, 16, 64)* & 0.153 & 0.386 & 0.582 & 0.744 \\ \hline
			(128, 2,048, 16, 64) & 0.059 & 0.199 & 0.348 & 0.488 \\ \hline
			(64, 4,096, 16, 64) 	& 0.033 & 0.107 & 0.190 & 0.286 \\ \hline
			(32, 8,192, 16, 64) 	& 0.052 & 0.068 & 0.108 & 0.160 \\ \hline
			(16, 16,384, 16, 64) 	& 0.058 & 0.048 & 0.066 & 0.089 \\ \hline
			(8, 32,768, 16, 64) 	& 0.077 & 0.036 & 0.050 & 0.056 \\ \hline
			(4, 65,536, 16, 64) 	& 0.063 & 0.049 & 0.037 & 0.048 \\ \hline
			(2, 131,072, 16, 64) & 0.070 & 0.040 & 0.054 & 0.044 \\ \hline
			(1, 262,144, 16, 64) 	& 0.059 & 0.043 & 0.044 & 0.053 \\ \hline
		\end{tabular}
		\caption{Average estimate error in third round of trials. $^*$ indicates trials retained from second round.}\label{ta:avgestimate3}
	\end{center}
\end{table}

\begin{table}[H]
	\begin{center}
		\begin{tabular}{| l | c | c | c | c |c | c | c | c |}
			\hline
			Relaxation Parameters & $n=4$	& $n=6$	& $n=8$	& $n=10$\\ \hline
			(512, 512, 16, 64) & 0 & 20 & 20 & 20 \\ \hline
			(256, 1,024, 16, 64)* & 0 & 0 & 20 & 20 \\ \hline
			(128, 2,048, 16, 64) & 0 & 0 & 0 & 5 \\ \hline
			(64, 4,096, 16, 64) & 0 & 0 & 0 & 0 \\ \hline
			(32, 8,192, 16, 64) & 0 & 0 & 0 & 0 \\ \hline
			(16, 16,384, 16, 64) & 0 & 0 & 0 & 0 \\ \hline
			(8, 32,768, 16, 64) & 0 & 0 & 0 & 0 \\ \hline
			(4, 65,536, 16, 64) & 0 & 0 & 0 & 0 \\ \hline
			(2, 131,072, 16, 64) & 0 & 0 & 0 & 0 \\ \hline
			(1, 262,144, 16, 64) & 0 & 0 & 0 & 0 \\ \hline
		\end{tabular}
		\caption{Incidence of estimation outside specified bounds in third round of trials. $^*$ indicates trials retained from second round.}\label{ta:misestimates3}
	\end{center}
\end{table}

\begin{table}[H]
	\begin{center}
		\begin{tabular}{| l | c | c | c | c |c | c | c | c |}
			\hline
			Relaxation Parameters & $n=4$	& $n=6$	& $n=8$	& $n=10$\\ \hline
			(512, 512, 16, 64) 	& 0 & 0 & 0 & 0 \\ \hline
			(256, 1,024, 16, 64)* & 0 & 0 & 0 & 0 \\ \hline
			(128, 2,048, 16, 64) 	& 0 & 0 & 0 & 0 \\ \hline
			(64, 4,096, 16, 64) 	& 0 & 0 & 0 & 0 \\ \hline
			(32, 8,192, 16, 64) 	& 0 & 0 & 0 & 0 \\ \hline
			(16, 16,384, 16, 64) & 0 & 0 & 0 & 0 \\ \hline
			(8, 32,768, 16, 64) 	& 0 & 0 & 0 & 0 \\ \hline
			(4, 65,536, 16, 64) 	& 0 & 0 & 0 & 0 \\ \hline
			(2, 131,072, 16, 64) & 0 & 0 & 0 & 0 \\ \hline
			(1, 262,144, 16, 64) 	& 0 & 0 & 0 & 0 \\ \hline
		\end{tabular}
		\caption{Incidence of failure of algorithm in third round of trials. $^*$ indicates trials retained from second round.}\label{ta:failures3}
	\end{center}
\end{table}

\begin{table}[H]
	\begin{center}
		\begin{tabular}{| l | c | c | c | c |c | c | c | c |}
			\hline
			Relaxation Parameters & $n=4$	& $n=6$	& $n=8$	& $n=10$\\ \hline
			(512, 512, 16, 64) & 49 & 1,126 & 11,240 & 52,571 \\ \hline
			(256, 1,024, 16, 64)* & 50 & 1,139 & 11,014 & 53,109 \\ \hline
			(128, 2,048, 16, 64) & 52 & 1,181 & 11,005 & 55,223 \\ \hline
			(64, 4,096, 16, 64) & 52 & 1,184 & 10,934 & 55,468 \\ \hline
			(32, 8,192, 16, 64) & 52 & 1,153 & 11,211 & 55,042 \\ \hline
			(16, 16,384, 16, 64) & 53 & 1,216 & 11,687 & 58,630 \\ \hline
			(8, 32,768, 16, 64) & 55 & 1,235 & 11,721 & 59,080 \\ \hline
			(4, 65,536, 16, 64) & 53 & 1,227 & 11,830 & 57,365 \\ \hline
			(2, 131,072, 16, 64) & 52 & 1,242 & 12,088 & 56,485 \\ \hline
			(1, 262,144, 16, 64) & 49 & 1,321 & 11,081 & 55,111 \\ \hline
		\end{tabular}
		\caption{Average runtime (in seconds) in third round of trials. $^*$ indicates trials retained from second round.}\label{ta:runtime3}
	\end{center}
\end{table}

\subsection{Round 4}
Finally, a fourth round of trials (again ont he same set of 80 randomly generated matrices) was run to evaluate the performance of the FPRAS under no relaxation of the phase sample requirement and high relaxation of phase resampling time. The upper limit of the phase resampling time relaxation factor was chosen to be greater than the phase resampling time for all matrix sizes used in the experiment, resulting in the collection of consecutive samples from the Markov chain. The fourth round consisted of 560 additional trials.

\begin{table}[H]
	\begin{center}
		\begin{tabular}{| c | c | c | c |}
			\hline
			Samples 	& Resampling Time 	& Samples 			& Resampling Time \\
			per Phase 	& in Phases 			& for Refinement 	& for Refinement \\ \hline
			1$^*$		& 262,144$^*$		& 16$^*$				& 64$^*$	 \\ \hline
			1			& 524,288			& 16					& 64 \\ \hline
			1			& 1,048,576			& 16					& 64 \\ \hline
			1			& 2,097,152			& 16					& 64 \\ \hline
			1			& 4,194,304			& 16					& 64 \\ \hline
			1			& 8,388,608			& 16					& 64 \\ \hline
			1			& 16,777,216			& 16					& 64 \\ \hline
			1			& 33,554,432			& 16					& 64 \\ \hline		
		\end{tabular}
		\caption{Relaxation factors for fourth round of trials. $^*$ indicates trials retained from third round.}\label{ta:relax3}
	\end{center}
\end{table}

The 560 new trials in the fourth round required 949.35 hours of CPU time.

\begin{table}[H]
	\begin{center}
		\begin{tabular}{| l | c | c | c | c |c | c | c | c |}
			\hline
			Relaxation Parameters & $n=4$	& $n=6$	& $n=8$	& $n=10$\\ \hline
			(1, 262,144, 16, 64)* 	& 0.059 & 0.043 & 0.044 & 0.053 \\ \hline
			(1, 524,288, 16, 64) & 0.057 & 0.047 & 0.054 & 0.056 \\ \hline
			(1, 1,048,576, 16, 64) & 0.056 & 0.045 & 0.030 & 0.042 \\ \hline
			(1, 2,097,152, 16, 64) & 0.056 & 0.062 & 0.051 & 0.050 \\ \hline
			(1, 4,194,304, 16, 64) & 0.059 & 0.044 & 0.043 & 0.046 \\ \hline
			(1, 8,388,608, 16, 64) & 0.064 & 0.065 & 0.040 & 0.049 \\ \hline
			(1, 16,777,216, 16, 64) & 0.077 & 0.055 & 0.047 & 0.043 \\ \hline
			(1, 33,554,432, 16, 64) & 0.055 & 0.046 & 0.048 & 0.046 \\ \hline
		\end{tabular}
		\caption{Average estimate error in fourth round of trials. $^*$ indicates trials retained from third round.}\label{ta:avgestimate4}
	\end{center}
\end{table}

\begin{table}[H]
	\begin{center}
		\begin{tabular}{| l | c | c | c | c |c | c | c | c |}
			\hline
			Relaxation Parameters & $n=4$	& $n=6$	& $n=8$	& $n=10$\\ \hline
			(1, 262,144, 16, 64)* & 0 & 0 & 0 & 0 \\ \hline
			(1, 524,288, 16, 64) & 0 & 0 & 0 & 0 \\ \hline
			(1, 1,048,576, 16, 64) & 0 & 0 & 0 & 0 \\ \hline
			(1, 2,097,152, 16, 64) & 0 & 0 & 0 & 0 \\ \hline
			(1, 4,194,304, 16, 64) & 0 & 0 & 0 & 0 \\ \hline
			(1, 8,388,608, 16, 64) & 0 & 0 & 0 & 0 \\ \hline
			(1, 16,777,216, 16, 64) & 0 & 0 & 0 & 0 \\ \hline
			(1, 33,554,432, 16, 64) & 0 & 0 & 0 & 0 \\ \hline
		\end{tabular}
		\caption{Incidence of estimation outside specified bounds in fourth round of trials. $^*$ indicates trials retained from third round.}\label{ta:misestimates4}
	\end{center}
\end{table}

\begin{table}[H]
	\begin{center}
		\begin{tabular}{| l | c | c | c | c |c | c | c | c |}
			\hline
			Relaxation Parameters & $n=4$	& $n=6$	& $n=8$	& $n=10$\\ \hline
			(1, 262,144, 16, 64)* & 0 & 0 & 0 & 0 \\ \hline
			(1, 524,288, 16, 64) & 0 & 0 & 0 & 0 \\ \hline
			(1, 1,048,576, 16, 64) & 0 & 0 & 0 & 0 \\ \hline
			(1, 2,097,152, 16, 64) & 0 & 0 & 0 & 0 \\ \hline
			(1, 4,194,304, 16, 64) & 0 & 0 & 0 & 0 \\ \hline
			(1, 8,388,608, 16, 64) & 0 & 0 & 0 & 0 \\ \hline
			(1, 16,777,216, 16, 64) & 0 & 0 & 0 & 0 \\ \hline
			(1, 33,554,432, 16, 64) & 0 & 0 & 0 & 0 \\ \hline
		\end{tabular}
		\caption{Incidence of failure of algorithm in fourth round of trials. $^*$ indicates trials retained from third round.}\label{ta:failures4}
	\end{center}
\end{table}

\begin{table}[H]
	\begin{center}
		\begin{tabular}{| l | c | c | c | c |c | c | c | c |}
			\hline
			Relaxation Parameters & $n=4$	& $n=6$	& $n=8$	& $n=10$\\ \hline
			(1, 262,144, 16, 64)* & 49 & 1,321 & 11,081 & 55,111 \\ \hline
			(1, 524,288, 16, 64) & 50 & 853 & 7,263 & 34,165 \\ \hline
			(1, 1,048,576, 16, 64) & 41 & 680 & 5,366 & 23,790 \\ \hline
			(1, 2,097,152, 16, 64) & 40 & 547 & 4,143 & 17,754 \\ \hline
			(1, 4,194,304, 16, 64) & 41 & 525 & 3,793 & 15,791 \\ \hline
			(1, 8,388,608, 16, 64) & 41 & 536 & 3,755 & 15,043 \\ \hline
			(1, 16,777,216, 16, 64) & 42 & 544 & 3,576 & 14,547 \\ \hline
			(1, 33,554,432, 16, 64) & 42 & 531 & 3,488 & 13,896 \\ \hline
		\end{tabular}
		\caption{Average runtime (in seconds) in fourth round of trials. $^*$ indicates trials retained from third round.}\label{ta:runtime4}
	\end{center}
\end{table}
